\documentclass[11pt,a4paper]{article}

\usepackage[utf8x]{inputenc}
\usepackage[T1]{fontenc}
\usepackage{color}
\usepackage{amssymb}
\usepackage{amsmath,amsfonts}
\usepackage{graphicx}
\usepackage{mathtools}
\usepackage{ragged2e}
\usepackage{psfrag}
\usepackage{relsize}
\usepackage{upgreek}
\usepackage{booktabs}

\usepackage{bm}
\usepackage{mathrsfs}
\usepackage{graphicx}
\usepackage{verbatim} 
\usepackage[english]{babel}
\usepackage[hypertexnames=false]{hyperref} 
\usepackage{url}
\hypersetup{
citecolor=red,
colorlinks=true,
filecolor=red,
linkcolor=blue,
linktocpage=true,
urlcolor=blue
} 
\usepackage[titletoc,toc,title]{appendix}
\numberwithin{equation}{section}
\usepackage{cleveref}
\usepackage{cite}
\usepackage{epsfig}
\usepackage{float}
\usepackage{enumitem}
\usepackage[font={footnotesize,it}]{caption}
\usepackage{cases}
\usepackage{tikz,empheq}

\newcommand{\spll}{{/\kern-0.2em/}}
\newcommand\trick[1]{}
\newcommand{\be}{\begin{equation}} 
\newcommand{\ee}{\end{equation}}
\newcommand{\eq}[1]{(\ref{#1})}
\newcommand{\bit}{\begin{itemize}}  \newcommand{\eit}{\end{itemize}}
\newcommand{\ben}{\begin{enumerate}}  \newcommand{\een}{\end{enumerate}}

\newcommand{\ket}[1]{|#1 \rangle}

\newcommand{\Tr}{\operatorname{Tr}}
\newcommand{\dd}{\mathrm d}
\newcommand{\ii}{\mathrm i}

\newcommand{\Tens}{\mathsf T}
\newcommand{\Qten}{\mathsf Q}
\newcommand{\Mten}{\mathsf M}
\newcommand{\Rp}{\mathcal R}
\newcommand{\Lam}{\Lambda}
\newcommand{\gN}{g_N}

\def\bea{\begin{eqnarray}}
\def\eea{\end{eqnarray}}
\def\la{\langle}
\def\ra{\rangle}

\def\Tr{{\rm Tr}}
\def\dd{\mathrm d}

\def\a{\alpha}

\def\d{\delta}  \def\D{\Delta}  

\def\ve{\varepsilon}

\def\k{\kappa}

\def\o{\omega}

\def\s{\sigma}  
\def\t{\tau}
 \def\Th{\Theta}

 \def\cH{{\cal H}} 
 \def\cK{{\cal K}} 
\def\cM{{\cal M}} \def\cN{{\cal N}} \def\cO{{\cal O}}
  \def\cR{{\cal R}}
  
\def\cV{{\cal V}}

\def\d{\delta}\def\D{\Delta}

 \def\del{\partial}

\def\uno{\mbox{1 \kern-.59em {\rm l}}}

\def\dag{\dagger}

\def\CC {\mathbb{C}}

\def\bcomment#1{}

 \def\ii{{\rm i}}

\def\IR{\relax{\rm I\kern-.18em R}}

\long\def\symbolfootnote[#1]#2{\begingroup%
\def\thefootnote{\fnsymbol{footnote}}\footnote[#1]{#2}\endgroup}

\newcommand{\nthu}{{\it Department of Physics, National Tsing-Hua University,
    Hsinchu 30013, Taiwan}}

\newcommand{\ctc}{{\it
Center for Theory-Computation-Data Science Research, 
National Tsing-Hua University, Hsinchu 30013, Taiwan}}

\newcommand{\ncts}{{\it Physics Division,
    National Center for Theoretical Sciences, Taipei 10617, Taiwan}}

\begin{document}
\begin{center}
\vspace{20pt}
  
\thispagestyle{empty}
              {\Large \bf Spectral Chaos from a Fluctuating Quantum Horizon}

\vspace{25pt}

Chong-Sun Chu

\vspace{0.2cm}              

\vspace{5pt}\nthu\\
\vspace{5pt}\ctc\\
\vspace{5pt}\ncts

\vspace{1cm}

\begin{abstract}

Black holes are conjectured to be maximally chaotic, with their
quantum spectra expected to exhibit universal random-matrix
correlations. This expectation, however, has so far rested largely on
holographic descriptions or on many-body models in which randomness is
introduced rather than derived from the microscopic degrees of freedom
of the horizon itself.  In the horizon matrix model, the fuzzy sphere
supports highly degenerate parton states whose multiplicity accounts for the
black-hole entropy. We show that quantum fluctuations of the horizon
geometry lift this degeneracy and generate an ensemble of one-parton
Hamiltonians. Both the operator structure and the rank-dependent
covariance of this ensemble are determined by the microscopic model,
rather than fixed by quenched couplings.

Despite this structured covariance, the resulting spectra exhibit GOE,
GUE, and GSE statistics dictated by their antiunitary symmetry
classes, with the adjacent-gap statistic becoming increasingly
self-averaging with matrix size. The model also realizes the universal
Pandey--Mehta orthogonal-to-unitary crossover, with the crossover
scale determined by the microscopic geometric covariance. A
rank-resolved variance analysis shows that the fluctuation strength
remains broadly distributed across tensor ranks up to the microscopic
cutoff, providing a microscopic basis for the observed Wigner--Dyson
universality and effectively nonlocal dynamics.  The antiunitary
symmetry acts directly on the fuzzy-sphere geometry, so the Dyson
class of a conditional Hamiltonian is determined by the antiunitary
symmetry of the underlying quantum geometry. This provides a
microscopic counterpart of the Stanford--Witten relation between
geometry and random-matrix universality in JT gravity. These results
establish a microscopic route from quantum horizon geometry to
black-hole spectral chaos.

\end{abstract}
\end{center}

\newpage
\setcounter{footnote}{0}

\tableofcontents

\section{Introduction}
\label{sec:intro}

One of the deepest questions in quantum black-hole physics concerns
the microscopic origin of black-hole microstates.  Beyond accounting
for the Bekenstein--Hawking entropy \cite{Bekenstein:1972tm,Bekenstein:1973ur},
these microscopic degrees of
freedom are expected to underlie the unitary Page curve of Hawking
radiation \cite{Page:1993wv,Almheiri:2020cfm},
provide a microscopic basis for the membrane paradigm
\cite{Damour:1978cg,MacDonald:1982zz,Price:1986yy,Thorne:1986iy}, and
exhibit the strong quantum-chaotic behavior characteristic of black
holes \cite{Sekino:2008he,Shenker:2013pqa,Shenker:2013yza,Cotler:2016fpe,
Saad:2018bqo}.  A satisfactory microscopic theory should therefore explain
not only the number of black-hole states, but also the dynamical
properties that distinguish a black hole from a generic many-body
quantum system.

The idea that black holes are highly chaotic quantum systems arose
initially from their remarkable scrambling properties.  Black holes
were argued to be fast scramblers \cite{Sekino:2008he}, dispersing
initially localized
quantum information over their microscopic degrees of freedom on a
time scale
\be
t_{\rm scr}\sim \beta\log S
\ee
Holographic studies subsequently gave a more
quantitative realization of this idea.  A perturbation of an AdS black
hole produces a highly boosted shock wave near the horizon, leading to
an exponentially growing sensitivity of later observables to the
perturbation \cite{Shenker:2013pqa,Shenker:2013yza}.  In thermal quantum
systems this growth can be diagnosed by out-of-time-order correlators
(OTOCs), whose Lyapunov exponent is constrained by the MSS chaos bound
\cite{Maldacena:2015waa},
\be
\lambda_L\leq \frac{2\pi}{\beta}.
\ee
The Sachdev--Ye--Kitaev (SYK) model
\cite{Sachdev:1992fk,Kitaev:2015}, which provides a
holographic model of near-extremal black-hole dynamics through its
relation to Jackiw--Teitelboim gravity
\cite{Kitaev:2017awl,Maldacena:2016hyu} on nearly
AdS$_2$ \cite{Kitaev:2015,Maldacena:2016upp},
remarkably saturates this bound in the strong-coupling limit
\cite{Kitaev:2015,Maldacena:2015waa}.
These developments have established a close connection
between black holes, fast scrambling, and maximal quantum chaos.

These results, however, characterize black-hole chaos primarily
through holographic observables or effective many-body descriptions.
They do not by themselves identify the microscopic spacetime degrees
of freedom whose dynamics is responsible for the chaos.  This raises a
more fundamental question: if a black-hole horizon possesses genuine
microscopic constituents, does their Hamiltonian itself exhibit the
spectral signatures of quantum chaos?  Put differently, can random
matrix universality \cite{Mehta2004RandomMatrices}
be seen directly in a microscopic model of the
horizon, rather than inferred from gravitational scattering, OTOCs, or
a holographically dual boundary theory?

The recently proposed microscopic horizon matrix model of
Ref.~\cite{Chu:2024qil} provides a setting in which this question can be
addressed directly.  In this construction, the black-hole horizon is
described by a fuzzy-sphere quantum geometry together with a population
of partons given by lowest-Landau-level (LLL) fermionic states
\cite{Chu:2026vzj,Chu:2026qom}.
The geometric and matter degrees of freedom are therefore
both represented within a finite-dimensional matrix quantum system.
Macroscopic quantities such as the black-hole mass $M$ and the
microstate entropy $S$ were derived from the microscopic construction,
and their dependence on the fuzzy-sphere radius $R$ was shown to
reproduce the corresponding Schwarzschild relations $M(R)$ and
$S(R)$.  The same framework also gives the horizon a dynamical
interpretation: when coupled consistently to gravity,
the construction admits a junction interpretation with an exterior
Schwarzschild region and, under the assumptions of  \cite{Chu:2026kqt}, a de
Sitter interior. Its microscopic degrees of freedom provide a
realization of the membrane paradigm
\cite{Damour:1978cg,MacDonald:1982zz,Price:1986yy,Thorne:1986iy}.
A nontrivial counting of fermionic zero modes uniquely fixes the
microscopic tunneling trajectory to be a monopole configuration,
providing a quantum-mechanical origin \cite{Chu:2026wyh} for Hawking radiation.

The important point for the present work is that the geometry is not a
fixed background.  Quantum fluctuations of the fuzzy sphere perturb
the parton Hamiltonian, so that the microscopic fermionic dynamics
directly probes fluctuations of the horizon geometry.  Once a
geometric configuration is specified, the fermions experience a
definite single-particle Hamiltonian. When the geometry is treated
quantum mechanically, its vacuum fluctuations induce an ensemble of
such conditional Hamiltonians.  This ensemble is highly structured:
its matrix elements are correlated by angular-momentum selection rules,
the fuzzy-sphere geometry, and the covariance of the underlying
bosonic vacuum fluctuations.  It is therefore not a conventional
random matrix ensemble imposed by hand.  Whether universal random
matrix behavior nevertheless emerges from this geometrically
constrained microscopic dynamics is a nontrivial question.

This question has a close historical precedent in nuclear physics.
Random-matrix theory was introduced by Wigner to describe the
complicated spectra of highly excited nuclei, where the detailed
many-body dynamics is too intricate for individual energy levels to be
predicted microscopically \cite{Wigner:1951,Wigner:1955lhh}.
The essential discovery
was that, although the microscopic nuclear Hamiltonian is highly
structured and certainly not literally random, its fine spectral
correlations are nevertheless governed by universal statistical laws
determined primarily by symmetry class\cite{Dyson:1962es,Dyson:1962eeu}.
Subsequent work established the
connection between these Wigner--Dyson statistics and quantum chaos
\cite{Bohigas:1983er}.  The present problem is similar in spirit, but
with an important difference: here the ensemble is generated directly
by quantum fluctuations of microscopic horizon geometry.  The question
is therefore whether such geometrically constrained fluctuations are
sufficient to drive the parton spectrum into the same universal
Wigner--Dyson regime.

This distinction is particularly important.  Random matrix behavior
in a microscopic horizon model would not merely demonstrate that one
can construct a random Hamiltonian with black-hole-like statistics.
Rather, it would show that universal spectral correlations arise from
the quantum fluctuations of the horizon's own spacetime degrees of
freedom.  In this sense the question is analogous to asking whether
the complicated microscopic geometric dynamics becomes universal at
the level of sufficiently fine spectral correlations, despite the
strongly structured covariance inherited from the underlying fuzzy
geometry.

The diagnostics employed in this study are based on random-matrix
theory (RMT) spectral statistics \cite{Mehta2004RandomMatrices}.
Their central principle is that,
after all exact symmetries have been resolved, the local spectral
correlations of a sufficiently non-integrable quantum Hamiltonian
approach those of the appropriate universal Dyson ensemble.
The relevant universality class is determined by the antiunitary
symmetries of the Hamiltonian: depending on their presence and
structure, one obtains orthogonal, unitary, or symplectic statistics.
RMT therefore provides a sharper test than spectral irregularity
alone.  In particular, level repulsion and local gap correlations probe
universal short-range spectral rigidity, while the spectral form factor
tests correlations over parametrically larger spectral scales.

We show that the microscopic fuzzy-sphere parton system exhibits these
characteristic RMT signatures of quantum chaos.  Its level-spacing
statistics display the expected level repulsion, its adjacent-gap
ratios approach the universal Dyson values, and its spectral form
factor develops the characteristic random-matrix correlations.
Moreover, by continuously breaking the relevant antiunitary symmetry,
we observe the corresponding Pandey--Mehta crossover between Dyson
classes \cite{Pandey:1982br,Mehta:1983ns}.
This crossover is especially significant because it tests
not merely agreement with an isolated RMT benchmark, but the universal
trajectory by which one symmetry class evolves into another.

A notable feature of the construction is that the effective randomness
is generated by quantum fluctuations of the horizon geometry rather
than by introducing quenched random couplings at the microscopic
level.  Conditional on a particular instantaneous geometry, the
parton Hamiltonian is completely deterministic.  The statistical
ensemble arises because the bosonic vacuum wave function samples a
distribution of geometric perturbations.  The resulting covariance is
therefore strongly constrained and differs conceptually from the
quenched disorder familiar from models such as SYK.  The emergence of
Dyson universality despite this structure indicates that the
microscopic spectral correlations become insensitive to much of the
detailed geometric covariance, while retaining precisely the symmetry
information required to determine the RMT class.

These results provide direct microscopic evidence that the
fuzzy-sphere parton system possesses the quantum-chaotic spectral
structure expected of a black-hole horizon.  Together with the
previous derivations of the Schwarzschild thermodynamic relations and
the microscopic membrane response, they add an independent dynamical
criterion supporting the identification of the system as a
microscopic quantum black-hole horizon.  More broadly, they suggest a
mechanism by which universal quantum chaos can emerge directly from
fluctuating microscopic spacetime geometry.

The paper is organized as follows.  Section~\ref{sec:fermion} derives
the fermion Hamiltonian in the perturbed fuzzy-sphere geometry and its
exact projection onto the two parton bands.  Section~\ref{sec:ensemble}
derives the vacuum-induced conditional ensemble, characterizes its
structured covariance, and determines its antiunitary symmetry class.
Section~\ref{sec:RMT} presents the numerical RMT diagnostics, including
the continuous Pandey--Mehta crossover.  Section~\ref{sec:physical}
compares the construction with quenched SYK disorder and develops its
physical interpretation in terms of instantaneous conditioning and
environmental decoherence.  Section~\ref{sec:discussion} summarizes
the main implications and limitations.  Appendix~A derives the
recoupling coefficient explicitly, while Appendix~B describes the
direct numerical construction.

\section{Fermionic Hamiltonian under Geometric Perturbations}
\label{sec:fermion}
\subsection{Microscopic horizon and normal-ordered fermions}

We start with the microscopic action of \cite{Chu:2024qil}
\be
S=\int \dd t\,\Tr\!\left[
\frac{1}{2a_0^2M_P}\dot X_a^2
+\frac{M_P}{N^2}\bigl([X_a,X_b]^2+4X_a^2\bigr)
+\ii\psi^\dagger\dot\psi
-\frac{a_2M_P}{N^2}\!:\psi^\dagger\sigma_aX_a\psi\!:
\right].
\label{eq:action}
\ee
The theory has the round fuzzy sphere
\be
X_a=J_a,\qquad [J_a,J_b]=\ii\epsilon_{abc}J_c,\qquad
J_aJ_a=J(J+1)\mathbf 1_N,\qquad J=\frac{N-1}{2}.
\label{eq:round}
\ee
as solution. The model has $N$ flavors of identical fundamental fermions.
Each fermion has the normal-ordered one-particle Hamiltonian
\be \label{hF}
H_F= \gN:\!\psi^\dagger K[X]\psi\!:
\qquad
K[X]:=\sigma^aX_a
\quad \gN :=\frac{a_2M_P}{N^2},
\ee
with the one particle matrix
\be
h_F := \gN K[X].
\ee
The full fermion
Hamiltonian is simply $\mathbf 1_{N_f}\otimes H_F$ with $N_f=N$.
In this paper, we are interested in the
one-particle level statistics and we will remove
this trivial flavor degeneracy and study the Hamiltonian \eq{hF}
for a single copy of flavor.

On the round sphere, denote the spinor spin $1/2$ by
$\bm S :=\bm\sigma/2$.
Couple the matrix angular momentum $J_a$ to $S_a$ by introducing
\be
   \cK_a=J_a+ S_a,
   \ee
then
\be
  K_0=\sigma^aJ_a=2\bm S\!\cdot\!\bm J =\bm \cK^2 - \bm J^2 -\bm S^2.
  \ee
  The  angular momentum $\bm \cK$ has the total quantum number $k = J-1/2$ or
  $J+1/2$, which we will denote as $k_\pm := J\pm 1/2$.
  Let $ \cH_k \simeq \CC^{2k+1}$ denotes a spin $k$ representation of $SU(2)$.
  The $2N$-dimensional one-flavor Hilbert space of the partons
  decomposes as
\be
\mathcal H_J\otimes\mathcal H_{1/2}
=\mathcal H_{k_+}\oplus\mathcal H_{k_-}
\ee
with 
\be
k_+ := J+\frac12=\frac N2,
\qquad
k_- :=J-\frac12=\frac{N-2}{2}.
\label{eq:bands}
\ee
The dimensions of $\mathcal H_{k_\pm}$ are, respectively, 
\be
d_+=2k_++1=N+1,
\qquad
d_-=2k_-+1=N-1,
\ee
and $K_0$ has the eigenvalues $J$ or $-(J+1)$ on $\mathcal H_{k_\pm}$.
It is useful to introduce the projectors
\be \label{proj}
P_+=\frac{K_0+J+1}{2J+1},
\qquad
P_-=\frac{J-K_0}{2J+1}.
\ee
then $K_0$ has the eigen-decomposition
\be \label{K0JJ}
K_0 = J P_+ - (J+1) P_-. 
\ee
Choose orthonormal eigenvectors $U_\alpha$ and $V_\rho$ in the positive and negative bands and expand
\be
\psi=\sum_{\alpha=1}^{N+1}U_\alpha u_\alpha
+\sum_{\rho=1}^{N-1}V_\rho d_\rho^\dagger.
\label{eq:ud}
\ee
Normal ordering fills the negative first-quantized band and
interprets its excitations as holes.  The round-sphere Hamiltonian becomes
\be
  H_0 = 
\varepsilon_u\sum_\alpha u_\alpha^\dagger u_\alpha
+\varepsilon_d\sum_\rho d_\rho^\dagger d_\rho,
\ee
with
\be \label{eud}
\varepsilon_u :=\gN J,
\qquad
\varepsilon_d:=\gN(J+1).
\ee
This shows that the parton energy levels are highly degenerate.
In Ref.~\cite{Chu:2024qil,Chu:2024edh},
it was shown that the high level of degeneracy of
these parton states precisely accounts for the
Bekenstein-Hawking entropy of black hole.
In Ref.~\cite{Chu:2026vzj,Chu:2026qom}, it was shown that these levels
are precisely the lowest Landau levels of
an intrinsic Berry magnetic field on the fuzzy sphere, and
that their transport under external electromagnetic fields gives rise to
a quantum membrane paradigm. This highly integrable structure of energy levels
is, however, very different from that expected for 
a quantum chaotic system. A natural
puzzle therefore arises: whether and how quantum chaos can nevertheless emerge
from this highly structured microscopic system.
The answer is affirmative and demonstrating this is the main purpose of
this paper.

It turns out quantum fluctuations of the fuzzy-sphere geometry play a
crucial role. Therefore, let us first give a review of the geometric fluctuation
over the fuzzy sphere. 
Following the notation of Ref.~\cite{Chu:2026dhx},
a general fluctuation of the fuzzy sphere can be decomposed as
\be \label{pert-X}
\delta X_a
=\sum_{L=1}^{N-1}\sum_{K=L-1}^{L+1}\sum_{Q=-K}^{K}
q_{LKQ}B_a^{LKQ},  
\ee
where $B_a^{LKQ}$ is a fuzzy vector harmonic
with orbital angular momentum $L$ and total angular momentum $K$, 
and is unit-normalized
\be
\sum_a\Tr\!\left[(B_a^{LKQ})^\dagger B_a^{L'K'Q'}\right]
=\delta_{LL'}\delta_{KK'}\delta_{QQ'}.
\ee
The $K=L$ modes describe gauge transformation instead of physical fluctuation
and so they are removed from \eq{pert-X}, leaving
only $K = L \pm 1$ to describe
the physical fluctuations.
Hermiticity implies
\be
(B_a^{LKQ})^\dagger=(-1)^QB_a^{LK,-Q},
\qquad
q_{LK,-Q}=(-1)^Qq_{LKQ}^*.
\label{eq:realityq}
\ee
In the spherical basis,
\be \label{Ba}
B_a^{LKQ}
=\sum_{M,r}\langle LM;1r|KQ\rangle T_{LM}(e_r)_a,
\ee
where $T_{LM}$ is a normalized fuzzy scalar harmonic.
See Ref.~\cite{Chu:2026dhx} for further details and conventions.

It is convenient to introduce a collective index $\a$, then the fluctuation
of coordinates and fluctuation of the Yukawa matrix are
\be
\d X_a = \sum_\a q_\a B_a^\a, \quad \a =(LKQ)
\ee
\be \label{yuk}
 K(\bm q)
 =
 K_0
 +\sum_{\alpha = (LKQ)}q_\alpha \cO_\alpha,
 \qquad \cO_\a := \s_a B_a^\a.
 \ee
Let us write in the $P_+\oplus P_-$ basis, then
\be
\delta K = 
\begin{pmatrix}
A&B\\ B^\dagger&D
\end{pmatrix},
\qquad
A:=P_+\delta K P_+,
\quad
B:=P_+\delta K P_-,
\quad
D:=P_-\delta K P_-.
\ee
The normal-ordered perturbed parton Hamiltonian reads
\be
\delta H_F=\gN\left[
 u^\dagger Au-d^\dagger D^Td
 +u^\dagger Bd^\dagger+dB^\dagger u
 \right] .
\ee
Here the $A$-term and $D$-term
\be \label{hud}
\delta h_u=\gN P_+\delta K P_+,
\qquad
\delta h_d=-\gN(P_-\delta K P_-)^T
\ee
are the physical intraband one-particle Hamiltonian.
Note that the minus transpose in \eq{hud} is physically essential
but is irrelevant for the study of spectral diagnostics. 
If we denote $\cM_- :=  P_- \d K P_-$, then $ -  \cM_-^T$ 
has eigenvalues equal to the negatives of those of $\cM_-$.
Adjacent spacings are merely reversed in order, and the
spectral form factor
$| \Tr e^{-it \d h_d}|^2 = |\Tr e^{-i t \gN \cM_-}|^2$ is unchanged.

The $B$ and $B^\dagger$ terms create or annihilate a $u$ particle
together with a $d$ hole.  They do not contribute to first-order
degenerate perturbation theory within a fixed one-$u$ or one-$d$
sector.  At second order, for an initial state in either sector, their
contribution to the physical intraband Hamiltonian has the schematic
form
\be
\delta h_s^{(2)}
\sim
\frac{(\gN)^2}{\Delta E_{\rm pair}}\,
P_s\delta K P_{\bar s}\delta K P_s
\sim
\frac{\gN}{N}\,
P_s\delta K P_{\bar s}\delta K P_s ,
\label{eq:secondorderB}
\ee
where $\bar s$ denotes the opposite band.  The energy denominator is
most directly obtained from the unnormal-ordered one-particle
spectrum.  Since the two unperturbed levels are $\epsilon_u$ and
$-\epsilon_d$, their separation is
\be
\Delta E_{\rm pair}
=
\epsilon_u-(-\epsilon_d)
=
\gN N .
\label{eq:pairgap}
\ee
After normal ordering, the same energy difference is interpreted as
the energy required to create a $u$ particle together with a $d$
hole.

The vacuum normalization of the geometric fluctuations used below is
most conveniently expressed in terms of the rescaled fluctuation
\be
 \widehat{\delta K}:=
 \frac{2}{\sqrt{a_0N}}\,\delta K .
\ee
For this matrix one has parametrically
\be
 \frac{1}{d_s}
 \left\langle
 \Tr\!\left(
 P_s\widehat{\delta K}P_{\bar s}
 \widehat{\delta K}P_s
 \right)
 \right\rangle
 =O(1),
 \qquad s=\pm .
 \label{interbandscale}
\ee
Equivalently, for the unrescaled fluctuation appearing in
\eqref{eq:secondorderB},
\be
 \frac{1}{d_s}
 \left\langle
 \Tr\!\left(
 P_s\delta K P_{\bar s}\delta K P_s
 \right)
 \right\rangle
 =O(N).
\ee
Thus the rms interband matrix scale is $O(\sqrt N)$.  It follows that
\be
 \delta h_s^{(2)}=O(\gN),
 \qquad
 \delta h_s^{(1)}=O(\gN\sqrt N),
\ee
and hence
\be
 \frac{\delta h_s^{(2)}}{\delta h_s^{(1)}}
 =O(N^{-1/2}).
\ee
Interband mixing is therefore parametrically suppressed at large $N$
and does not affect the leading spectral statistics within either
parton band.

\subsection{Exact projection into a fixed fermion band}

The perturbed Hamiltonian \eq{hud} is given by a sum of products of
the magnitude $q_{LKQ}$ of fluctuation and the projected Yukawa operator
$P_k\mathcal O_{LKQ}P_k$ for the two bands $k = k_+, k_-$ of partons.
To proceed, let us first work out explicitly the operator part.
In appendix A, we show that
$\cO_{LKQ}$ is a rank-$K$ tensor under the total rotation
acting on the spinor Hilbert space, and 
the Wigner--Eckart theorem gives the following compact and explicit expression:
\be \label{WE}
P_k\mathcal O_{LKQ}P_k
=\Rp_{LK}^{(k)}\Tens_{KQ}^{(k)},
\ee
where the reduced coefficient $\Rp_{LK}^{(k)}$ is given by
\be \label{R9j}
\Rp_{LK}^{(k)}
=(2k+1)\sqrt{6(2L+1)}
\begin{Bmatrix}
J&\frac12&k\\
J&\frac12&k\\
L&1&K
\end{Bmatrix}.
\ee
The  tensor matrix $\Tens_{KQ}^{(k)}$
is  Hilbert-Schmidt  normalized in the sense
\be
\Tr\!\left[(\Tens_{KQ}^{(k)})^\dagger\Tens_{K'Q'}^{(k)}\right]
=\delta_{KK'}\delta_{QQ'}
\ee
and
has the matrix elements 
\be\label{Tmatrix}
\langle k,m'|\Tens_{KQ}^{(k)}|k,m\rangle
=(-1)^{k-m'}\sqrt{2K+1}
\begin{pmatrix}
k&K&k\\
-m'&Q&m
\end{pmatrix}.
\ee
Note that \eq{Tmatrix} gives the selection rule 
\be
  m'=m+Q.
\label{eq:selection}
\ee
and hence a rank-$K$ multiplet occupies at most the first $K$
off-diagonals on each side of the matrix.
As a result, we obtain
\be
h_u = \ve_u+ \gN \sum_{LKQ} q_{LKQ} \Rp_{LK}^{(+)}\Tens_{KQ}^{(k_+)},\qquad
h_d = \ve_d - \gN \sum_{LKQ} q_{LKQ} \Rp_{LK}^{(-)}\Tens_{KQ}^{(k_-) T}
\ee

For the numerical analysis, 
it is useful to note that  $9j$ symbol can be evaluated completely.
For the $K=L+1$ branch,   we have for the $u$ band ($k=k_+=N/2$) and
the negative band ($k=k_-=(N-2)/2$),
\be \label{R+}
\Rp^{(+)}_{L,L+1}
=+\sqrt{
  \frac{(L+1)(N+L+1)(N+L+2)}{(2L+3)N^2}}, \quad
\Rp^{(-)}_{L,L+1}
=-\sqrt{
\frac{(L+1)(N-L-2)(N-L-1)}{(2L+3)N^2}}.
\ee
Similarly, for the $K=L-1$ branch, we have
\be \label{R-}
\Rp^{(+)}_{L,L-1}
=-\sqrt{
\frac{L(N-L)(N-L+1)}{(2L-1)N^2}}, \qquad
\Rp^{(-)}_{L,L-1} =+\sqrt{
\frac{L(N+L-1)(N+L)}{(2L-1)N^2}}.
\ee

\section{Ensemble from Quantum Geometric Fluctuations}
\label{sec:ensemble}

\subsection{Emergence of ensemble of Hamiltonians}

Next let us consider the amplitude $q_{LKQ}$ of fluctuation.
For numerical analysis later on, it is convenient
to employ a real basis so that
the corresponding fluctuation amplitudes are independent.
For example, for every $Q>0$, we can use the real basis of fuzzy harmonics:
\begin{align} \label{real-b}
\Qten_{K,0}^{(k)}&: =\Tens_{K0}^{(k)},\qquad
\Qten_{K,cQ}^{(k)}
:=\frac{\Tens_{KQ}^{(k)}+(-1)^Q\Tens_{K,-Q}^{(k)}}{\sqrt2},\qquad
\Qten_{K,sQ}^{(k)}
 :=\frac{\Tens_{KQ}^{(k)}-(-1)^Q\Tens_{K,-Q}^{(k)}}{\ii\sqrt2}.
\end{align}
Each matrix is Hermitian and the set is Hilbert-Schmidt orthonormal.
As a result, the corresponding fluctuation amplitudes $q_{LK,A}$ are real
and the reality condition imposes no further relation among the real amplitudes.
Here $A = \{0, cQ,sQ \}$, $Q =1, 2, \cdots, K$ are $(2K+1)$ labels for the
real basis \eq{real-b}.

Let us consider
a unit-normalized physical mode, the quadratic effective Lagrangian is
\be\label{L_q}
L_{LK,A}=\frac{1}{2a_0^2M_P}
\left[
\dot q_{LK,A}^{\,2}
-\Omega_0^2\Lam_{LK}^{\rm eff}(N)q_{LK,A}^{\,2}
\right],
\qquad
\Omega_0=\frac{2a_0M_P}{N}.
\ee
As shown in Ref.~\cite{Chu:2026dhx}, the finite-$N$ curvatures 
\be \label{eq:Leff}
 \Lam_{LK}^{\rm eff}(N)=\Lam_{LK}^{\rm cl}+\frac{\pi N}{6}C_{LK}(N),
 \qquad a_0=\frac\pi3,
\ee
with the classical physical branches
\be
 \Lam_{L,L-1}^{\rm cl}=L(L+3),
 \qquad
 \Lam_{L,L+1}^{\rm cl}=(L+1)(L-2).
\ee
are positive for large and finite $N$.
The classically unstable $(1,2)$ and classically flat
$(2,3)$ modes are included with their positive finite-$N$ one-loop curvatures.
For the values of $N$ analyzed below, every physical $K>0$ mode has positive
$\Lam^{\rm eff}$.
The vacuum wavefunction of the
simple harmonic oscillator \eq{L_q} therefore induces a Gaussian 
probability distribution for the geometric coordinate $q_{LK,A}$. With
\be \label{oLK}
M_q :=\frac{1}{a_0^2M_P},
\qquad
\omega_{LK} :=\Omega_0\sqrt{\Lam_{LK}^{\rm eff}(N)},
\ee
the ground-state variance is
\be\label{eq:qvariance}
\langle q_{LK,A}^2\rangle
=\frac{1}{2M_q\omega_{LK}}
=\frac{a_0N}{4\sqrt{\Lam_{LK}^{\rm eff}(N)}}.
\ee
We can therefore write
\be \label{q-xi}
q_{LK,A}=\frac{\sqrt{a_0N}}{2}
[\Lam_{LK}^{\rm eff}(N)]^{-1/4}\xi_{LK,A},
\ee
where $\xi_{LK,A}$ has unit variance
\be
\langle \xi_{LK,A}^2\rangle =1.
\ee

For spectral statistics, the constant round-sphere energies and
multiplicative common scale factor are irrelevant. Therefore, after
removing a common scale factor of $\Gamma_N : = \gN \sqrt{a_0N}/2$,
the spectral statistics of the horizon partons system
is determined by energy levels of the Hamiltonians
\be \label{HuQ}
\mathcal H_s
:= 
\sum_{L,K=L\pm1,A}
\frac{\Rp_{LK}^{(s)}\,\xi_{LK,A}}
{[\Lam_{LK}^{\rm eff}(N)]^{1/4}}
\Mten_{KA}^{(k_s)},
\ee
where
\be \label{Mten}
\Mten^{(k_s)}_{KA}:=
\begin{cases}
\Qten^{(k_+)}_{KA}, & k_s=k_+,\\[2mm]
-\left(\Qten^{(k_-)}_{KA}\right)^T, & k_s=k_-.
\end{cases}
\ee
Since transposition and multiplication by an overall minus sign preserve
the Hilbert--Schmidt inner product, the matrices $\Mten^{(k_s)}_{K,A}$ obey
\be
\Tr\!\left(
\Mten^{(k_s)}_{K,A}\Mten^{(k_s)}_{K',A'}
\right)
=
\delta_{KK'}\delta_{AA'}.
\ee

At this point, it is interesting to note that the normalized normal
coordinates $\{\xi_{LK,A}\}$ acquire a natural statistical description.
To see this, recall that the fuzzy sphere is taken to be in its
ground state.  For the geometric sector this means that it is in the
quantum state
\be \label{vac-geom}
 |0_{\rm geom}\rangle
 =
 \int d\bm q\,
 \Psi_0(\bm q)\,
 |\bm q\rangle ,
 \ee
rather than in a state of sharply defined geometry $\bm q=0$.  The
ground state is centered on the round fuzzy sphere,
\be
 \langle q_\alpha\rangle_0=0, \qquad \a = (LK,A),
\ee
but possesses finite zero-point fluctuations,
\be
 \langle q_\alpha^2\rangle_0\neq0.
\ee
Thus the round fuzzy sphere represents the mean geometry of the quantum
ground state,
whereas the state itself has support on a distribution of nearby
geometric configurations. Since the quantum geometry is not measured,
these configurations are weighted by
the ordinary Born probability. In terms of $\xi_\a$, it is 
\be \label{P-Psi}
 P_0(\bm \xi)=|\Psi_0(\bm \xi)|^2=
\prod_{\a}
\frac{1}{\sqrt{2\pi}}
\exp\left(-\frac{\xi_\a^2}{2}\right),
\ee
where $\xi_\a$ is related to $q_\a$ by \eq{q-xi}.
As a result,  one obtains naturally an ensemble of parton Hamiltonians,
\be \label{ens-H}
\{ \cH_u(\{\xi\}),\cH_d(\{\xi\})\}_{\{\xi\sim\cN(0,1)\}}.
\ee
It is important to distinguish the Born probability distribution from
an incoherent statistical mixture.  In the geometric configuration
basis, the ground-state density matrix is
\be
 \rho_0(\bm q, \bm q')
 =
 \Psi_0(\bm q)\Psi_0^*(\bm q'),
\ee
and therefore
\be
 P_0(\bm q)
 :=
 \rho_0(\bm q,\bm q)
 =
 |\Psi_0(\bm q)|^2 .
\ee
The off-diagonal elements $\rho_0(\bm q,\bm q')$, $\bm q\neq \bm q'$,
are in general
nonzero and encode the quantum coherence between different geometric
configurations.  Thus the use
of $P_0(\bm q)$ at this stage does not assume
that the geometric state has decohered into a classical statistical
mixture.
Nevertheless, for any quantity that is diagonal in the geometric
configuration basis, its quantum expectation value depends only on the
diagonal element $P_0(\bm q)$ of the density matrix.  In particular,
a spectral quantity $F_s = F[H_s(\bm q)]$ has the expectation value
\be \label{F-vev}
 \langle F_s\rangle
 =
 \int d\bm q\,P_0(\bm q)\,F[H_s(\bm q)] .
\ee
The sampling of geometric configurations employed below is therefore simply a
Monte Carlo representation of this Born-weighted expectation value.
We note that, by itself, this sampling does not replace the coherent geometric
state by a classical statistical ensemble.  Such an effective ensemble
interpretation may arise only after decoherence, as discussed in
Sec.~\ref{sec:physical}.

We remark that the black hole actually has a temperature.
This follows from the fact,
although this point was not emphasized in \cite{Chu:2024qil}, that the system
possesses a thermodynamic temperature determined by $T = \del
E/\del S$ from the relations of $S =S(R)$ and $E =E(R)$ obtained for the
fuzzy sphere parton system. 
This agrees precisely, as required for consistency,
with the Hawking temperature $T_H = 1/4\pi R$
obtained independently from the quantum-mechanical analysis of Hawking
radiation. For purposes of thermal equilibrium,
the reduced geometric sector may be described by the thermal density
operator
\be
 \rho_{\rm geom}
 =
 \frac{1}{Z_{\rm geom}} e^{-H_{\rm geom}/T_H},
 \label{rho-geom}
 \ee
 where 
\be
 H_{\rm geom} =
 \sum_\alpha
 \hbar\omega_\alpha
 \left(a_\alpha^\dagger a_\alpha+\frac12\right)
 \label{Hgeomosc}
 \ee
describes the excitation of the quantum geometric fluctuations of the horizon. 
Here the oscillation frequency is given by \eq{oLK}.
In the geometric configuration basis, the density matrix  induces a
probability  distribution 
\be
 P_T(\bm q) :=
 \langle \bm q|\rho_{\rm geom}|\bm q\rangle 
 \label{Pthermal}
 \ee
 for its horizon geometries.
 We use $\bm q$ for the physical geometric coordinates and
 $\bm \xi$ for the corresponding standardized unit-variance coordinates.
 The explicit form of the distribution
 is controlled by the mean thermal occupation.
For a harmonic mode $\alpha$,
\be \label{nbarthermal}
 \bar n_\alpha
 =
 \frac{1}{e^{\Xi_\alpha}-1}, \qquad
 \Xi_{LK} :=
 \frac{\hbar\omega_{LK}}{T_H} .
 \ee
For a harmonic mode, the  thermal distribution is
\be
 P_{T,\alpha}(q_\alpha)
 =
 \langle q_\alpha|\rho_\alpha|q_\alpha\rangle
 =
 \frac{1}{\sqrt{2\pi\sigma_{\alpha,T}^2}}
 \exp\left[
 -\frac{q_\alpha^2}{2\sigma_{\alpha,T}^2}
 \right],
 \label{PTq}
\ee
where
\be
 \sigma_{\alpha,T}^2
 =
 \sigma_{\alpha,0}^2
 \coth\left(\frac{\Xi_\alpha}{2}\right),
 \qquad
 \sigma_{\alpha,0}^2
 =
 \langle0|q_\alpha^2|0\rangle .
 \label{thermalwidth}
\ee
For the Hawking temperature $ T_H=\frac{1}{4\pi N\ell_P}$, we have
\be
 \Xi_{LK}
 =
 8\pi a_0M_P\ell_P
 \sqrt{\Lambda_{LK}^{\rm eff}} .
 \label{xLK}
 \ee
 and the thermal behaviour
 is controlled by the effective curvature itself. Now recall 
 the results of \cite{Chu:2026dhx}:  The exceptional
 rotational invariant scale mode has $\Lambda_{10}^{\rm eff}\to4$,
 so $\Xi_{10}=O(1)$. Numerically, taking $a_0=\pi/3$ and
$M_P\ell_P \simeq 1$,  the exceptional homogeneous scale mode
has $\Xi_{10} \simeq 52.6$ and hence $\bar n_{10}\sim 10^{-23}$.
 A non-saturating fixed-$L$ physical mode receives
an $O(N)$ one-loop stiffness and therefore has $\Xi_{LK}=O(\sqrt N)$.
The mesoscopic region $L\sim\sqrt N$ has $\Lambda^{\rm eff}=O(N)$ and
the same $\Xi_{LK} = O(\sqrt N)$ scaling.  At fixed $x=L/N$, the classical
curvature is $O(N^2)$, so $\Xi_{LK}=O(N)$.  Hence
\be
\Xi_{LK}\sim
\begin{cases}
O(1), & (L,K)=(1,0),\\ O(\sqrt N), & L\ \text{fixed and
  non-saturating},\\ O(\sqrt N), & L\sim\sqrt N,\\ O(N), & L=xN.
\end{cases}
\label{eq:XiScaling}
\ee
Therefore,
\be
P_{T,\alpha}(q_\alpha)
 =
 |\psi_{0,\alpha}(q_\alpha)|^2
 \left[1+O(e^{-\Xi_\alpha})\right]
 \ee
 for the quantum geometric sector.
Although  the black hole is thermal,  the
geometric modes that generate the random-matrix level correlations are
effectively frozen into their quantum zero-point distributions
and take the fuzzy horizon to be at
the quantum ground state \eq{vac-geom}. We remark this is for the
geometric sector only. Thermality of the horizon has important effects on
the dynamics of the partons, for example, in giving rise to  the Ohm's
conductance of the horizon for the parton electric current.

\subsection{Anti-unitary symmetry and Dyson classification}
\label{sec:symmetry}
Random-matrix theory classifies quantum Hamiltonians according to
their symmetry under antiunitary transformations. The basic Gaussian
ensembles are generated from a probability measure of the form
\be
P(H) dH  \propto e^{-\frac{\beta}{2}\Tr H^2}dH.
\ee
Depending on the  presence or absence of
antiunitary symmetries, different classes of matrices are sampled.
If  there is no antiunitary  symmetry $\Th$ satisfying $ \Th H\Th^{-1}=H$,
no reality condition is imposed and the generic Hamiltonian  is
complex Hermitian.   The corresponding ensemble is the Gaussian
unitary ensemble (GUE), with Dyson index  $\beta=2$. If an antiunitary
symmetry exists with $\Theta^2=1$, 
a basis can be chosen in which $H$ is real symmetric.
The corresponding ensemble is the
Gaussian orthogonal ensemble (GOE), with $\beta=1$.
Finally, if an antiunitary symmetry exists with $\Theta^2=-1$,
the spectrum exhibits Kramers degeneracy and
the corresponding ensemble is the Gaussian
symplectic ensemble (GSE), with $\beta=4$.

When random-matrix universality is realized, the
Dyson index  controls the universal short-distance repulsion
of neighboring eigenvalues,
\be
P(s)\sim s^\beta,\qquad s\rightarrow0,
\ee
so that GOE, GUE, and GSE exhibit respectively linear, quadratic, and
quartic level repulsion.
The antiunitary symmetry therefore determines which Dyson
class is expected if random-matrix universality emerges. Whether
the spectrum actually exhibits such universality is a separate
dynamical question that must be established from its spectral
statistics.

\paragraph{Time-reversal parity of a Hermitian rank-$K$ component.}

In standard quantum mechanics, there is a time-reversal operator given by
\be
\Th=e^{-\ii\pi\cK_y}\mathcal C,
\ee
where $\bm{\cK}$ is the total angular momentum and $\mathcal C$
denotes the complex conjugation.  In a spin-$k$ representation $\cH_k$ for the
partons, its action is
\be
 \Th\ket{k,m}=(-1)^{k-m}\ket{k,-m}
\ee
and hence
\be
 \Th^2=(-1)^{2k}.
 \label{eq:theta2}
\ee
Using Eq.~\eqref{eq:bands}, both projected sectors obey
\be
 \Th^2=(-1)^N.
 \label{eq:thetaN}
\ee
Thus, for even $N$, the projected spins are integer and
$\Th^2=+1$, so a time-reversal-invariant Hamiltonian would belong to
the orthogonal class (GOE). For odd $N$, the projected spins are
half-integer and $\Th^2=-1$, so a time-reversal-invariant Hamiltonian
would belong to the symplectic class (GSE), with Kramers degeneracy.

Let us return to our parton system. 
It is natural to decompose the parton Hamiltonian
\eq{HuQ} into a sum of irreducible tensors of rank $K$. For a fixed $s$, we
have
\be
\cH_s = \sum_K \cH_K^{(s)},
\ee
where
\be
\cH_K^{(s)} := \sum_A Y_{KA}^{(s)}\Mten_{KA}^{(k_s)}, \qquad
Y_{KA}^{(s)}
:=\sum_{L=K\pm1}
\frac{\Rp_{LK}^{(s)}}{[\Lam_{LK}^{\rm eff}(N)]^{1/4}}\,
\xi_{LK,A}.
\ee
We now show that each rank-$K$ tensor component of
the Hamiltonian has  a definite
parity under the antiunitary time reversal symmetry
\be \label{HKparity}
\Th \cH^{(s)}_K \Th^{-1} = (-1)^K \cH^{(s)}_K.
\ee
To see this, note that
the irreducible tensor $\Tens_{KQ}^{(k)}$ transforms as
\be
 \Th\Tens_{KQ}^{(k)}\Th^{-1}
 =(-1)^{K-Q}\Tens_{K,-Q}^{(k)}.
 \label{eq:TRtensor}
\ee
This can be proven by recalling that in the standard $\cK_z$ basis,
\be
\left[\Tens_{KQ}^{(k)}\right]_{m'm}
=
(-1)^{k-m'}\sqrt{2K+1}
\begin{pmatrix}
k & K & k\\
-m' & Q & m
\end{pmatrix}.
\ee
With the usual Condon--Shortley convention, the $3j$ symbols are real,
and hence
$ \mathcal C\,\Tens_{KQ}^{(k)}\,\mathcal C^{-1} =
\Tens_{KQ}^{(k)}$,
therefore
\be
\Th\Tens_{KQ}^{(k)}\Th^{-1}
=
e^{-\ii\pi\cK_y}
\Tens_{KQ}^{(k)}
e^{\ii\pi\cK_y}.
\ee
For an irreducible tensor operator one has
\be
U(R)\Tens_{KQ}^{(k)}U(R)^{-1}
=
\sum_{Q'}
D^{(K)}_{Q'Q}(R)\Tens_{KQ'}^{(k)},
\ee
where
\be
D^{(j)}_{m'm}(\alpha,\beta,\gamma)
=
\langle j,m'|R(\alpha,\beta,\gamma)|j,m\rangle .
\ee
is the Wigner $D$-matrix. For the standard $z$--$y$--$z$ Euler-angle convention,
\be
D^{(j)}_{m'm}(\alpha,\beta,\gamma)
=
e^{-im'\alpha}\,
d^{(j)}_{m'm}(\beta)\,
e^{-im\gamma}, \qquad 
d^{(j)}_{m'm}(\beta)
=
\langle j,m'|
e^{-i\beta J_y}
|j,m\rangle, 
\ee
where $d^{(j)}_{m'm}(\beta)$ is the Wigner small-$d$ matrix,
corresponding to a rotation by
$\beta$ about the $y$ axis.
For a rotation by $\pi$ about the $y$ axis, we thus have
\be
D^{(K)}_{Q'Q}\big(R_y(\pi)\big)
=
(-1)^{K-Q}\delta_{Q',-Q},
\ee
which proves Eq.~\eqref{eq:TRtensor}.
Thus time reversal reverses the magnetic component $Q\to -Q$, and gives
a phase factor $(-1)^{K-Q}$ appropriate to a rank-$K$ spherical
tensor. When the $Q$ and $-Q$ components are combined into
the real Hermitian basis $\Mten^{(k)}_{KA}$, the relation \eq{eq:TRtensor}
becomes simply 
\be
\Th\Mten_{KA}^{(k)}\Th^{-1}
=
(-1)^K \Mten_{KA}^{(k)}.
\ee
As a result,
every rank-$K$ Hamiltonian component has a definite time-reversal parity
\eq{HKparity}.

Therefore, a generic realization containing any nonzero odd-$K$
component breaks the microscopic time-reversal symmetry  and removes
the antiunitary constraint responsible for the GOE/GSE classification.
In the absence of any additional antiunitary symmetry, if
random-matrix universality emerges the resulting spectrum is therefore
expected to belong to the GUE Dyson class. As we show below, the
numerical spectra indeed exhibit GUE statistics.

To understand more clearly how the symmetry structure controls the
spectral statistics, 
it is useful to consider separately the time reversal symmetric
ensemble obtained by retaining
only the even-$K$ tensor components,
\be
\cH_{\rm even}^{(s)}
:=
\sum_{\substack{K\ {\rm even}}}
\cH_K^{(s)}.
\ee
For this ensemble, the expected Dyson class depends on the parity of
$N$. For even $N$, the projected spins are integer and
$\Th^2=+1$, so the even-$K$ ensemble is expected to belong to the GOE
class. For odd $N$, the projected spins are half-integer and
$\Th^2=-1$, so the even-$K$ ensemble is expected to belong to the GSE
class, with Kramers-degenerate energy levels.
We may therefore summarize the Dyson classes expected from the
microscopic symmetry analysis, provided random-matrix universality
emerges, as
\be
\begin{array}{c|c}
\text{ensemble} & \text{expected Dyson class} \\ \hline
\text{all $K$} & \mathrm{GUE}\\
\text{even-$K$, even $N$} & \mathrm{GOE}\\
\text{even-$K$, odd $N$} & \mathrm{GSE}
\end{array}
\ee
We emphasize, however, that the symmetry analysis determines only the
Dyson class that would be compatible with random-matrix universality;
whether such universality is actually realized is a separate dynamical
question. We test this directly below for the ensemble of
fluctuation-induced Hamiltonians 
using their spectral statistics.

\subsection{Structured covariance}

To characterize the ensemble, it is useful to introduce the covariance 
$\left\langle
(\mathcal H_s)_{ij}
(\mathcal H_s)_{kl}^{*}
\right\rangle $
for the matrix elements of the parton Hamiltonian. It is easy to obtain
\be
\left\langle
(\mathcal H_s)_{ij}
(\mathcal H_s)_{kl}^{*}
\right\rangle
=
\sum_{K,A}w_{K,s}^{2}
(\Mten_{KA}^{(k_s)})_{ij}
(\Mten_{KA}^{(k_s)})_{kl}^{*},
\label{eq:matrixcov}
\ee
where
\be\label{wKs}
w_{K,s}^2
:=\sum_{L=K\pm1}
\frac{|\Rp_{LK}^{(s)}|^2}
{\sqrt{\Lam_{LK}^{\rm eff}(N)}}
\ee
is the covariance
of the rank-$K$ tensor amplitudes $Y_{KA}^{(s)}$
\be
\left\langle Y_{KA}^{(s)}Y_{K'A'}^{(s)*}\right\rangle
=w_{K,s}^2\, \delta_{KK'}\delta_{AA'} .
\ee
Note that $w_{K,s}^{2}$ receives contributions from
the two physical orbital branches
\be \label{2orbits}
(L,K)=(K-1,K),\qquad (K+1,K)
\ee
that lead to the same rank $K$. Since the corresponding
Gaussian normal coordinates are independent, their
variances add in \eq{wKs}.
We emphasize that  the covariance \eqref{eq:matrixcov} of
our  fluctuation induced ensemble carries nontrivial $K$-dependent weights
$w_K^2$, so different fuzzy-sphere tensor channels contribute with
different fluctuation strengths. The resulting  structured covariance,
fixed by the microscopic fuzzy-sphere tensor decomposition,
is therefore markedly different from the
unitarity or orthogonal invariant 
covariance underlying the standard GUE or GOE random-matrix ensembles.
It is thus nontrivial if, as we shall demonstrate in the next section,
the fluctuation-induced ensembles
nevertheless exhibit the corresponding Dyson spectral statistics
and the associated random-matrix signatures of quantum chaos.

\section{Random-Matrix Universality of the Horizon Partons}
\label{sec:RMT}

In this section, we present our results for the spectral statistics of the
\emph{quantum-geometric-fluctuation-induced ensemble of Hamiltonians}
\eq{ens-H}. 
We  use the ground-state probability distribution \eq{P-Psi} to generate
independent realizations of the geometric fluctuations and hence of the
corresponding perturbed Hamiltonian.  Each realization is then
diagonalized to
obtain its perturbed energy levels, from which we compute standard spectral
correlation diagnostics, including the adjacent-gap ratio,
nearest-neighbor
spacing distribution and the spectral form factor.
Remarkably, the resulting spectral correlations display random-matrix
universality with high accuracy.  In addition, the fluctuation-induced
ensemble reproduces the expected Dyson symmetry classification and, upon
continuously breaking the relevant antiunitary symmetry, exhibits the
characteristic Pandey--Mehta crossover between symmetry classes
\cite{Pandey:1982br,Mehta:1983ns}.

Moreover, the random-matrix behavior is a typical property of the
quantum-geometric ensemble rather than an effect produced only after
averaging over many realizations. To explain ``typicality'' in this
context,   it is useful to elucidate what constitutes a ``typical''
vacuum configuration.
After the $K=0$ identity component is omitted, the complete two-branch
physical fluctuation space contains
\be
 D_N=\sum_{L=1}^{N-1}(2L+3)+\sum_{L=2}^{N-1}(2L-1)
 =2N^2-3
 \label{eq:Dcount}
\ee
independent standardized real coordinates.  The joint Gaussian density is
maximal at $\boldsymbol\xi=0$, but the radial probability measure is
\be
 p(R)\propto R^{D_N-1}e^{-R^2/2},
 \qquad R^2=\sum_\alpha\xi_\alpha^2,
 \label{eq:radialmeasure}
\ee
and is concentrated at $R\simeq\sqrt{D_N}$.  Thus the exactly round
configuration is highly atypical in the large-dimensional vacuum measure.
A {\it typical vacuum configuration}
contains many simultaneous nonzero geometric
fluctuations and defines a nontrivial parton Hamiltonian whose spectral
chaos we wish to test.

We find that a single typical realization already
exhibits the characteristic local Wigner--Dyson level correlations.  In
particular, the realization-to-realization fluctuations of the
matrix-averaged adjacent-gap ratio decrease as $N$ increases.
The data are consistent with
\be
 \sigma_{\bar r}\sim N^{-1/2},
\ee
suggesting a convergence in probability 
\be
 \bar r(\{\xi\})=\langle\bar r\rangle_{\rm RMT}
             +O_p(N^{-1/2}).
\ee
Thus local spectral statistics exhibits $N^{-1/2}$ self-averaging:
almost every typical quantum geometry produces the same universal
random-matrix behavior, while ensemble averaging primarily improves the
statistical precision with which this universality is resolved.

The use of the vacuum measure as a statistical measure is dictated by
Eq.~\eqref{P-Psi} when the geometric fluctuations are left unobserved.
Repeated sampling from this measure is, in addition, extremely useful
numerically.  
At fixed rank $N$, a single parton band contains only $O(N)$
adjacent-gap ratios, whereas independent vacuum realizations at the same
rank provide a much larger statistical sample of the corresponding
fixed-$N$ spectral statistics.  This is also computationally advantageous.
The cost of constructing the parton Hamiltonian grows polynomially with
$N$: for example, the number of angular-momentum components in the fuzzy
harmonic basis grows as $O(N^2)$, while diagonalization of a dense
matrix has the usual $O(N^3)$ cost.  It is therefore computationally
more economical to improve the statistical precision by
sampling many independent vacuum configurations at
moderate $N$ than to increase $N$
solely in order to obtain more levels from a single spectrum.
The vacuum measure thus induces the ensemble
\be
 \left\{
 \mathcal H_u(\boldsymbol\xi),
 \mathcal H_d(\boldsymbol\xi)
 \right\}_{\boldsymbol\xi\sim\mathcal N(0,1)},
\ee
which provides an efficient Monte Carlo representation of the
Born-weighted distribution and allows the spectral correlations of
typical  Hamiltonians to be estimated with much higher statistical
precision.

\subsection{Single-realization typicality and self-averaging}
\label{sec:singlecopy}

We begin without averaging over geometric configurations.  For one fixed draw
of  $\bm \xi$, Eq.~\eqref{HuQ} is a single matrix whose diagonalization gives
one spectrum, from which the
adjacent-gap ratios can be obtained from  averaging over the set of levels.
This is the standard realization-by-realization logic of spectral quantum
chaos.
\begin{table}[ht]
\caption{Single-realization adjacent-gap test. ``Full'' retains all physical
$K>0$ tensor ranks and is expected to realize GUE. ``Even $K$'' retains only
time-reversal-even ranks; for these even values of $N$ the expected class is
GOE.
Parentheses give the deviation from the appropriate reference mean in
units of its single-realization standard deviation. For
$D_N^{-1}\sum_\alpha\xi_\alpha^2$, the reference is the Gaussian
vacuum distribution; for $\bar r$, the reference is the matched
finite-dimensional RMT ensemble.}
\label{tab:singlecopy}
\centering
\begin{tabular}{rcccc}
\toprule
$N$ & $D_N$
& $D_N^{-1}\sum\xi_\alpha^2$
& $\bar r_{\rm full}$
& $\bar r_{\rm even}$\\
\midrule
100 & $19{,}997$
    & $1.0108\;(1.08\sigma)$
    & $0.5585\;(-2.00\sigma)$
    & $0.5157\;(-0.66\sigma)$\\
500 & $499{,}997$
    & $1.0010\;(0.50\sigma)$
    & $0.5995\;(+0.02\sigma)$
    & $0.5277\;(-0.29\sigma)$\\
\bottomrule
\end{tabular}
\end{table}

Table~\ref{tab:singlecopy} shows two representative large-$N$ vacuum
configurations.  For $N=100$ the complete finite-$N$ curvatures are taken
from the exact one-loop stability scan.
For $N=500$ where a direct evaluation of the complete finite-$N$
one-loop curvature is computationally expensive,
we use large-$N$ form established in Ref.~\cite{Chu:2026dhx} to
reconstruct the curvatures. In the bulk, we employ the fixed-$x=L/N$
extrapolation
\be
 C_{LK}(500,x)
 =1.6\,C_{LK}(200,x)-0.6\,C_{LK}(100,x),
\ee
while the low-$L$ modes are treated by fixed-$L$ expansions, with
analytic formulas for the exceptional $(1,0)$, $(1,2)$, and $(2,3)$
modes.  For the extreme UV regime
$x>0.9$, the $N=100$ edge profile is continued using the
$N=500-N=100$ shift determined at $x=0.9$.
The normalized radius
$D_N^{-1}\sum_\alpha\xi_\alpha^2$ is very close to unity in both cases,
confirming that the configurations are typical rather than special points
near $\boldsymbol\xi=0$.
At $N=100$ a single band contains only about one hundred levels, so a
single-spectrum mean $\bar r$ still has substantial finite-sample
fluctuations.  Both values are compatible with the corresponding matched
finite-dimensional RMT single-spectrum distributions.  At $N=500$ the
self-averaging is much sharper: pooling the $u$ and $d$ spectral estimators
gives
\be
 \bar r_{\rm full}=0.59952,
 \qquad
 \bar r_{\rm even}=0.52774,
 \label{eq:r500single}
\ee
which compare favorably with the
finite-dimensional GUE and GOE means near $0.599$ and $0.531$.
We remark that,
as explained above,
for $N=500$ we use the large-$N$ curvature reconstruction of
Ref.~\cite{Chu:2026dhx}, rather than a complete finite-$N$ curvature
calculation.  The individual reconstructed curvatures therefore carry
extrapolation uncertainty.  We use this large-$N$ example only to
illustrate the increasing self-averaging of the local spectral statistic,
rather than claiming a  precision prediction at $N=500$.

A stronger finite-size test is to compute one value of $\bar r$ for each
vacuum realization and compare the resulting distribution of $\bar r$
with that obtained from matched finite-dimensional RMT.  For a common
geometric draw $a$, we pool the $u$ and $d$ results together 
\be\label{pool}
 \bar r_a^{\rm pool}
 :=\frac{(d_u-2)\bar r_{u,a}+(d_d-2)\bar r_{d,a}}
 {(d_u-2)+(d_d-2)},
 \qquad d_u=N+1,\quad d_d=N-1
\ee
with weights equal to the numbers of gap
ratios. Note that the two spectra are never concatenated.
For each value of $N$, we generate independent vacuum realizations and
determine the realization-level adjacent-gap statistic
$\bar r_a^{\rm pool}$.  We compare its mean and realization-to-realization
width with those obtained from the corresponding matched
finite-dimensional GUE ensemble.  The results are shown in
Table~\ref{tab:selfavg}.
\begin{table}[ht]
\centering
\caption{Realization-level adjacent-gap statistics for the full conditional
horizon Hamiltonian and the corresponding matched finite-dimensional GUE
reference ensemble. For each value of $N$, both ensembles contain 12,000
realizations. The quoted width $\sigma$ is the standard deviation of the
realization-level pooled adjacent-gap statistic $\bar r$.}
\begin{tabular}{c@{\qquad}cc@{\qquad}cc}
\toprule
$N$ & $\langle\bar r\rangle_{\rm hor}$ & $\sigma_{\rm hor}$ & $\langle\bar r\rangle_{\rm GUE}$ & $\sigma_{\rm GUE}$ \\
\midrule
12 & 0.597957 & 0.061742 & 0.597647 & 0.062846 \\
15 & 0.598072 & 0.055089 & 0.597686 & 0.055366 \\
20 & 0.597740 & 0.047045 & 0.597837 & 0.047051 \\
25 & 0.598655 & 0.042019 & 0.598376 & 0.041251 \\
30 & 0.598458 & 0.037592 & 0.597703 & 0.037642 \\
45 & 0.598701 & 0.030680 & 0.598916 & 0.030757 \\
50 & 0.598810 & 0.028814 & 0.598587 & 0.029270 \\
60 & 0.599275 & 0.026395 & 0.598377 & 0.026491 \\
\bottomrule
\end{tabular}
\label{tab:selfavg}
\end{table}

\begin{figure}[t]
\centering
\includegraphics[width=0.68\textwidth]
{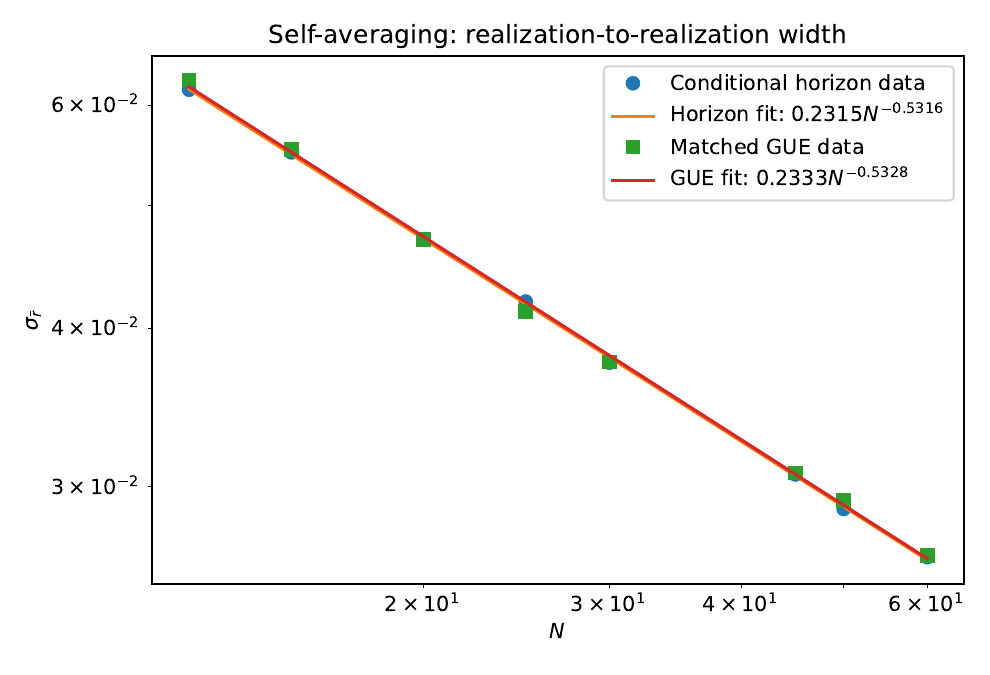}
\caption{Realization-to-realization width of the pooled adjacent-gap statistic.
The conditional horizon ensemble and matched finite-dimensional GUE show the
same nearly $N^{-1/2}$ self-averaging law.}
\label{fig:selfavgsigma}
\end{figure}

The realization-to-realization widths are well described by
\be \label{self-r}
 \sigma^{\rm hor}_{\bar r}(N)
 = 0.2315\,N^{-0.5316},
 \qquad
 \sigma^{\rm GUE}_{\bar r}(N)
 = 0.2333\,N^{-0.5328},
\ee
with $R^2_{\rm hor}=0.99973$ and $R^2_{\rm GUE}=0.99887$, respectively.
The curves shown in Fig.~\ref{fig:selfavgsigma} are these
unconstrained power-law fits.
As a separate check, fixing the exponent to $-1/2$ gives
\be
 \sigma^{\rm hor}_{\bar r}\simeq\frac{0.2104}{\sqrt N},
 \qquad
 \sigma^{\rm GUE}_{\bar r}\simeq\frac{0.2114}{\sqrt N}.
 \ee

\begin{figure}[t]
\centering
\includegraphics[width=0.68\textwidth]
{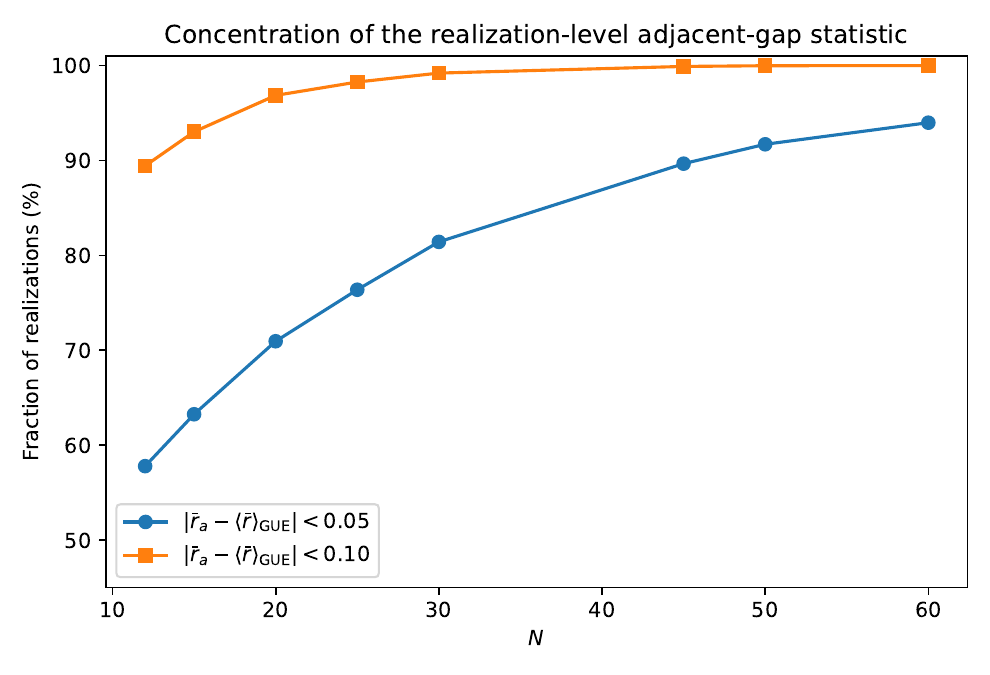}
\caption{Fraction of individual conditional Hamiltonians whose
realization-level $\bar r$ lies within a fixed tolerance of the matched GUE
mean.  A single typical realization becomes increasingly representative as
$N$ grows.}
\label{fig:selfavgtypicality}
\end{figure}

We can also examine the concentration of the realization-level
statistics.  The fractions of individual horizon realizations satisfying
$|\bar r_a-\langle\bar r\rangle_{\rm GUE}|<0.05$ increase from
$57.8\%$ at $N=12$ to $94.0\%$ at $N=60$, while the corresponding
fractions within $0.10$ increase from $89.4\%$ to $99.99\%$.
Figure~\ref{fig:selfavgtypicality} displays this concentration.
Self-averaging is  therefore  a stronger statement than agreement of the
ensemble averages:
\be
  P_{\rm vac}\!\left(\bar r[H_{\rm p}(\mathbf q)]\right)
 \simeq P_{\rm GUE}^{(d)}(\bar r),
 \qquad
 \operatorname{Var}_{\rm vac}(\bar r)\to0.
 \label{eq:selfavgdistribution}
\ee

The spectral self-averaging observed here is analogous to
thermodynamic self-averaging.  If an intensive statistical-mechanical
observable is the average of many weakly correlated local contributions,
then away from criticality its relative variance typically scales as the
inverse volume.  In the present case we find
\be
 \frac{\operatorname{Var}(\bar r)}
      {\langle\bar r\rangle^2}
 \simeq \frac{0.125}{N},
\ee
so that the realization-to-realization fluctuations of $\bar r$ decrease
as $N^{-1/2}$.  Thus a typical large-$N$ Hamiltonian becomes increasingly
representative of the ensemble value, in direct analogy with an ordinary
self-averaging thermodynamic observable.
We remark however that as in statistical mechanics, such self-averaging is not
automatic for every observable or every system.  Critical, glassy, or
strongly correlated systems can violate the simple inverse-size law.
Likewise, the present result applies specifically to the local spectral
statistic $\bar r$ and need not extend to all spectral observables.

\subsection{Numerical setup}

Self-averaging explains why one sufficiently large spectrum is representative;
many-realization sampling remains the most efficient way to obtain precision
finite-$N$ correlation functions. 
We analyze the following 8 values of 
\be
 N=12,15,20,25,30,45,50,60.
\ee
For every $N$, the exact finite-$N$ bosonic one-loop curvatures
$\Lambda_{LK}^{\rm eff}(N)$ are supplied by the stability calculation of
Ref.~\cite{Chu:2026dhx}.  The relevant parton Hamiltonian
matrices are assembled directly from
Eq.~\eqref{HuQ}, with independent real Gaussian coordinates assigned to every
physical branch $(L,K)$ and every Hermitian tensor component $A$.  We
generate $12,000$ geometric realizations for each $N$.
The $u$ and $d$ bands are diagonalized and analyzed separately.
Spectral
statistics such as the adjacent-gap ratio are first computed
separately within each band
and are combined only afterward with the appropriate statistical
weights.  The two sets of eigenvalues are never merged into a single
ordered spectrum.  For odd $N$ in the even-$K$ ensemble, time-reversal
symmetry implies exact Kramers degeneracy; each Kramers pair is
replaced by its mean, and the resulting distinct levels are used
for determining the spectral statistics.
To account for finite $N$ effects, 
every obtained data set is compared with an independently generated
finite-dimensional GOE, GUE, or GSE benchmark of
the corresponding one-particle Hilbert-space dimension.
The adjacent-gap ratio requires no unfolding.  As for $P(s)$
and the spectral form factor, each fixed-$N$ band and symmetry
sector is unfolded separately with an ensemble-mean odd-quintic staircase,
\be
 \overline{{\cal N}}_5(E)=c_0+c_1u+c_3u^3+c_5u^5,
 \label{eq:unfold}
\ee
defined over its fitted spectral support.  Beyond the lower and upper
endpoints of this support, the unfolding map is continued by the tangent
line to the fitted staircase at the corresponding endpoint, rather than
by extrapolating the quintic polynomial.  This tangent-linear
continuation preserves monotonicity of the unfolding map near the
spectral edges.  The matched finite-dimensional RMT spectra are processed
with the same unfolding prescription, with the staircase fit performed
independently for each RMT reference ensemble.  A numerical check confirms
that all fitted unfolding maps remain monotonic over the spectra used in
the analysis, so that the ordering of neighboring levels is preserved.
Implementation details and additional validation checks are given in
Appendix~\ref{app:numerics}.

We emphasize
the use of many realizations in the numerical calculation should not be
interpreted as implying that random-matrix behavior arises only after
ensemble averaging.  The self-averaging property \eq{self-r}
shows that the local RMT spectral correlations are already present in a
typical conditional Hamiltonian, with the realization-to-realization
fluctuations decreasing as $N$ increases.  In particular, the
Wigner--Dyson level repulsion found above is therefore a property of
individual typical spectra rather than an artifact of ensemble
averaging. Many-realization averaging, however,
serves a different purpose: it provides an
efficient way to determine finite-$N$ spectral correlation functions
with high statistical precision.

\subsection{Adjacent-gap ratio}
\label{sec:r}

For an ordered spectrum
\be
E_1<E_2<\cdots<E_d,
\ee
we define
\be
\delta_n=E_{n+1}-E_n,
\qquad
r_n=
\frac{\min(\delta_n,\delta_{n+1})}
     {\max(\delta_n,\delta_{n+1})},
\ee
and the realization mean
\be
\bar r
=
\frac{1}{d-2}
\sum_{n=1}^{d-2} r_n .
\label{eq:rbar}
\ee
The adjacent-gap ratio is computed directly from the raw spectrum and
requires no unfolding.
The $u$ and $d$ bands are analyzed separately.  If their retained
dimensions are $d_u$ and $d_d$, the corresponding realization means
$\bar r_u$ and $\bar r_d$ are combined using \eq{pool}.
We remark that pooling leaves the spectra themselves
separate.  In particular, the $u$- and $d$-band eigenvalues are never
combined into a single ordered spectrum. For odd $N$ in the even-$K$
ensemble, $d_u$ and $d_d$ denote the numbers of distinct levels remaining
after one eigenvalue from each Kramers pair has been retained.
Table~\ref{tab:rsummary} summarizes the corresponding pooled results for the
symmetry-resolved ensembles.
\begin{table}[ht]
\centering
\caption{Adjacent-gap ratios of the branch-resolved horizon ensemble from the
same 12,000 geometric realizations used in Table~\ref{tab:selfavg}. The full
physical ensemble is compared with matched finite-dimensional GUE. The even-$K$
ensemble is compared with GOE for even $N$ and Kramers-reduced GSE for odd $N$.
The shown uncertainties denote one standard error of the mean.}
\begin{tabular}{c@{\qquad}cc@{\qquad}ccc}
\toprule
$N$ & full horizon & matched GUE & even-$K$ class & even-$K$ horizon & matched RMT \\
\midrule
12 & 0.597957(564) & 0.597647(574) & GOE & 0.528981(617) & 0.528520(625) \\
15 & 0.598072(503) & 0.597686(505) & GSE & 0.671619(660) & 0.671351(663) \\
20 & 0.597740(429) & 0.597837(430) & GOE & 0.530131(463) & 0.529265(464) \\
25 & 0.598655(384) & 0.598376(377) & GSE & 0.672403(486) & 0.671256(483) \\
30 & 0.598458(343) & 0.597703(344) & GOE & 0.529165(375) & 0.529898(375) \\
45 & 0.598701(280) & 0.598916(281) & GSE & 0.672332(349) & 0.672248(353) \\
50 & 0.598810(263) & 0.598587(267) & GOE & 0.530045(288) & 0.530356(285) \\
60 & 0.599275(241) & 0.598377(242) & GOE & 0.530231(260) & 0.530481(262) \\
\bottomrule
\end{tabular}
\label{tab:rsummary}
\end{table}

The full-ensemble adjacent-gap ratios agree with the matched GUE for all eight
ranks. The largest individual normalized discrepancy is $2.63$ standard errors,
and the combined test gives
\be
 \chi^2_{\rm full}= 10.71\quad (8\ {\rm points}),\qquad p\simeq0.22.
\ee
There is therefore no statistically significant departure from the matched
finite-dimensional GUE sequence. The class-matched even-$K$ data give
\be
\chi^2_{\rm even}=7.89\quad (8\ {\rm points}), \qquad p\simeq 0.44.
\ee
We remark that the even-$K$ calculation offers a stringent internal
consistency check since the same microscopic construction has to
realize two different antiunitary symmetry classes depending only on
the parity of $N$.
Here the quoted
discrepancies are normalized by the combined standard error of the horizon and
matched-RMT ensemble means
\be
z_N :=
\frac{
  \langle \bar r\rangle_{\rm hor}(N) - \langle \bar r\rangle_{\rm RMT}(N)}
     {\sqrt{ {\rm SE}_{\rm hor}^2(N) + {\rm SE}_{\rm RMT}^2(N) }},
\ee
so that the combined statistic is
\be
\chi^2=\sum_N z_N^2.
\ee
For all values of $N$ studied, the adjacent-gap
ratios agree with the corresponding finite-dimensional Dyson-class
benchmarks: GUE for the full ensemble, and GOE or Kramers-reduced GSE
for the even-$K$ ensemble according to the parity of $N$. The
simultaneous realization of all three Dyson classes within the same
microscopic construction, with the class determined solely by the
antiunitary symmetry of the underlying geometric fluctuations, provides
a nontrivial check of the random-matrix universality of the
conditional horizon Hamiltonian.

\subsection{Nearest-neighbor spacing distribution}
\label{sec:spacing}

\begin{figure}[t]
\centering
\includegraphics[width=0.33\textwidth]{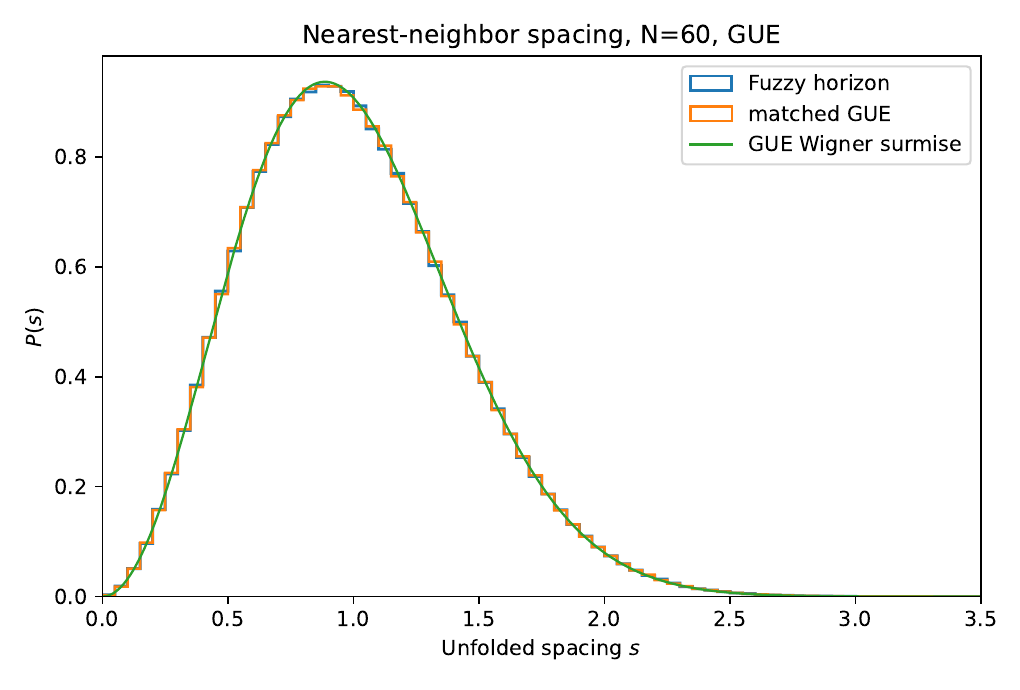}\hfill
\includegraphics[width=0.34\textwidth]{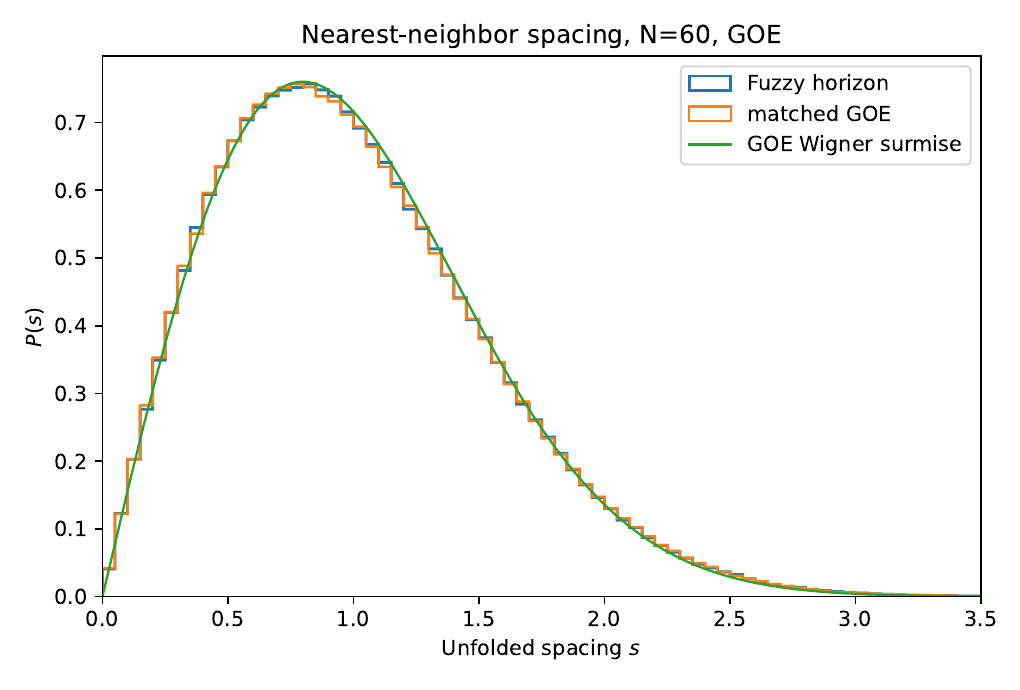}\hfill
\includegraphics[width=0.33\textwidth]{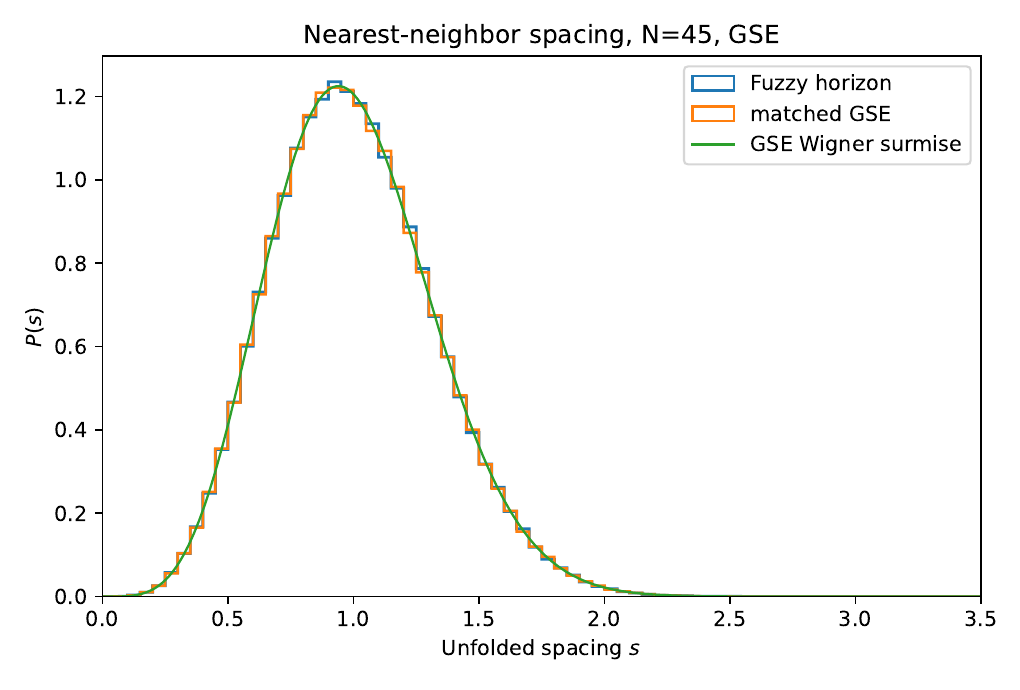}
\caption{Representative nearest-neighbor spacing distributions.  Left: full
$N=60$ ensemble versus GUE.  Center: even-$K$, $N=60$ versus GOE.  Right:
Kramers-reduced even-$K$, $N=45$ versus GSE.}
\label{fig:spacing}
\end{figure}

The adjacent-gap ratio compresses local correlations into one number.
A stronger test is provided by the full unfolded nearest-neighbor spacing
distribution $P(s)$. 
For each $N$, the spectrum is unfolded to $\{x_n\}$, so that
the nearest-neighbor level spacing
\be
 s_n :=x_{n+1}-x_n,
\ee
has  unit mean $\la s\ra =1$. 
A selection of our results for $P(s)$ is shown in Figure \ref{fig:spacing}.
Our result for the level-spacing distribution $P(s)$
exhibits the expected small-spacing level repulsion
$P(s)\sim s^\beta$ with $\beta=1,2,4$ for the
GOE, GUE, and GSE respectively. More quantitatively, we compare the
numerical spacing distributions directly with the corresponding
finite-dimensional RMT results. Across the full and symmetry-preserving
ensembles at all eight ranks, the two-sample Kolmogorov--Smirnov test gives
\be
D_{\rm KS}\le 2.340\times10^{-3}.
\label{eq}
\ee
Thus the microscopic and corresponding finite-dimensional RMT
cumulative spacing distributions differ by at most $0.234\%$ over the
full spacing range, showing close quantitative agreement.

\subsection{Connected spectral form factor}
\label{sec:sff}

The connected spectral form factor (SFF)
probes the same universal two-level correlations in Fourier space
and extends the test beyond the nearest-neighbor repulsion. 
For each unfolded realization,  define the spectral trace
\be
 Z_a(\tau)=\sum_{n=1}^{d}e^{-2\pi\ii\tau x_{n,a}}
\ee
The ensemble-averaged spectral form factor and its connected part are
\be
 K(\tau)=\frac1d\langle |Z_a(\tau)|^2\rangle_a,
 \qquad
 K_c(\tau)=\frac1d\left[
 \langle |Z_a(\tau)|^2\rangle_a
 -|\langle Z_a(\tau)\rangle|_a^2
 \right],
 \ee
 where $\la \cdot \ra_a$ denotes the average over vacuum realizations.
Here the connected subtraction removes the smooth mean-density contribution and
isolates the universal two-level correlations.
We note that the unfolded variable $x_n$ has unit mean level spacing,
$\langle x_{n+1}-x_n\rangle=1$, although the total unfolded spectral
range grows with the number of levels.  With the convention
$Z(\tau)=\sum_n e^{-2\pi i\tau x_n}$, the conjugate variable $\tau$
is therefore a dimensionless spectral time measured in units of the
Heisenberg time associated with the local mean level spacing.  In
particular, the Heisenberg scale occurs at $\tau=O(1)$ for every $N$.
Consequently the connected SFFs for different matrix sizes can be
displayed over the same range of $\tau$ without any additional
$N$-dependent rescaling. Finite-$N$ effects appear in the shape of the
curves rather than in an overall shift of the Heisenberg scale.

\begin{figure}[t]
\centering
\includegraphics[width=0.32\textwidth]{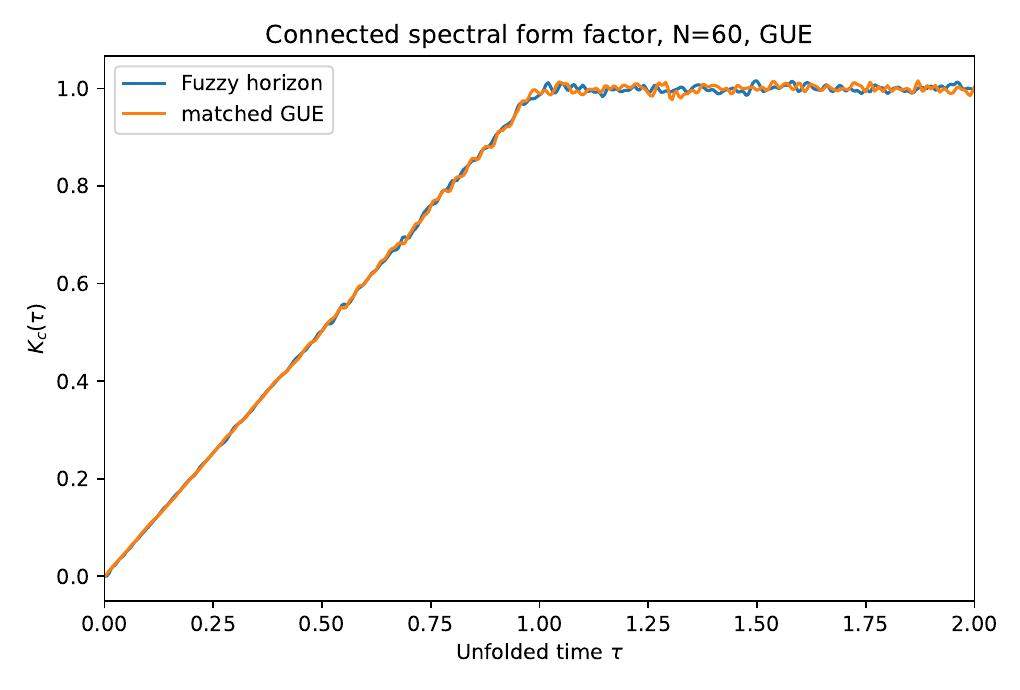}\hfill
\includegraphics[width=0.32\textwidth]{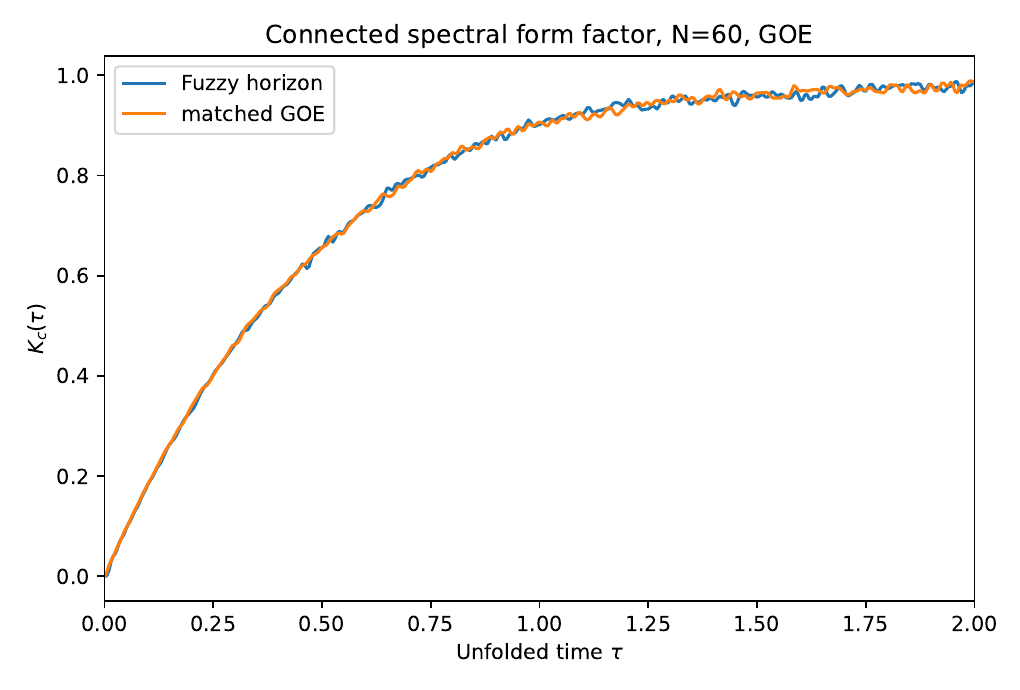}\hfill
\includegraphics[width=0.32\textwidth]{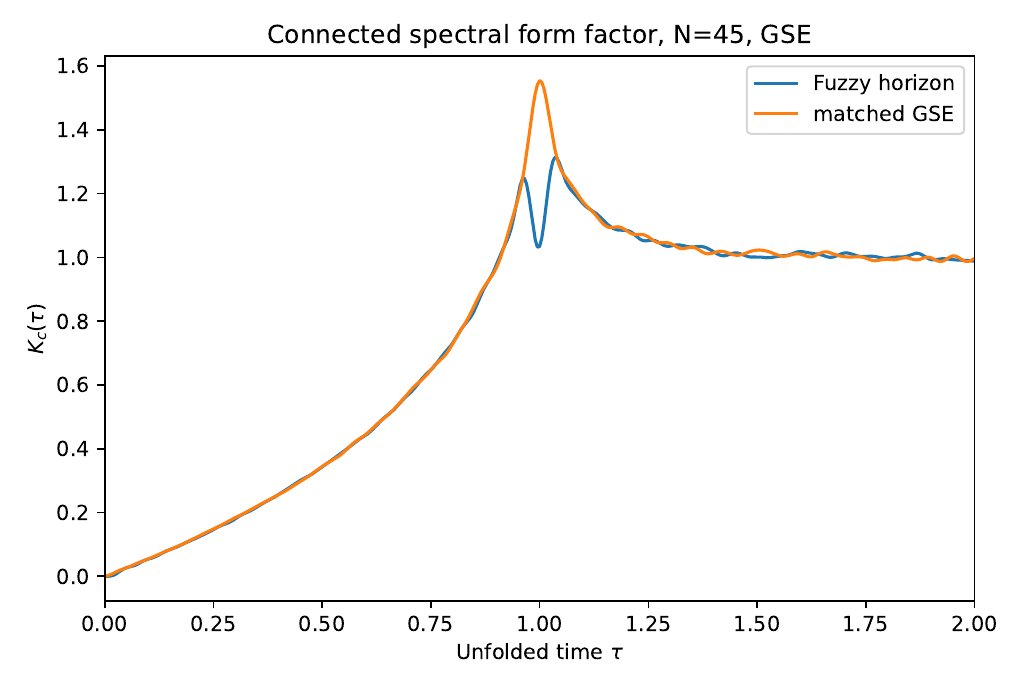}
\caption{Representative connected spectral form factors.  Left: full
$N=60$ GUE.  Center: even-$K$, $N=60$ GOE.  Right: Kramers-reduced even-$K$,
$N=45$ GSE.}
\label{fig:sff}
\end{figure}

Our results are shown in Figure \ref{fig:sff}.
We note that the connected SFFs in the GUE and GOE symmetry classes agree
closely with their matched finite-dimensional RMT benchmarks,
reproducing
the approximately linear GUE ramp and the characteristic curved GOE ramp,
respectively, followed in both cases by the late-time plateau.
 After
Kramers-pair reduction, the GSE sector likewise exhibits the expected
symplectic ramp and plateau. However there are larger finite-$N$ deviations
near the Heisenberg time scale that decrease systematically with
increasing $N$:
\be
 0.13471\ (N=15)\ \longrightarrow\
 0.10271\ (N=25)\ \longrightarrow\
 0.07492\ (N=45),
 \label{eq:GSEsff}
\ee
where the quoted values are the connected-SFF
root-mean-square error over the comparison
window.

It is interesting to ask if the SFF is also self-averaging.
To test it,  consider the coefficient of variation  defined by
\be
 {\rm CV}_K(\tau)
 :=
 \frac{\sigma_{K_a}(\tau)}
      {\langle K_a(\tau)\rangle_a} =
\frac{
\sqrt{
\left\langle K_a(\tau)^2\right\rangle_a
-
\left\langle K_a(\tau)\right\rangle_a^2
}
}{
\left\langle K_a (\tau)\right\rangle_a
},
\ee
where $\sigma_{K_a}(\tau)$ is the standard deviation over vacuum
realizations.  Thus ${\rm CV}_K(\tau)\ll1$ would signal self-averaging,
whereas ${\rm CV}_K(\tau)=O(1)$ indicates realization-to-realization
fluctuations comparable to the ensemble mean.
Using the same ensemble unfolding,  we obtain the results for  ${\rm CV}_K(\tau)$
for the two parton bands:
\be
\begin{array}{c|cccccccc}
\tau&0.025&0.05&0.10&0.20&0.50&1.00&1.50&2.00\\ \hline
u&0.066&0.92&0.96&1.00&0.99&0.97&0.98&0.96\\
d&0.056&0.94&0.98&0.99&0.99&0.97&0.96&0.97
\end{array}
\label{eq:SFFcvN20}
\ee
Thus the full SFF is relatively smooth only at the earliest spectral times.
By $\tau=0.05$ the  fluctuations are already
of order the mean and remain so throughout the ramp and plateau. For
example, at $\tau=1$ the upper band gives
\be
 \langle K\rangle_{\rm hor}=1.0023,
 \qquad \sigma_K^{\rm hor}=0.9767,
\ee
while the matched $21\times21$ GUE gives $1.0048$ and $0.9819$, respectively.
The late-time non-self-averaging is therefore itself part of the GUE
behavior, rather than a failure of universality.  Averaging a single
realization over the window $1.5\le\tau\le2$ reduces the relative width to
approximately $0.22$ in each band, illustrating how spectral/time smoothing
exposes the underlying ramp and plateau.  The same single-realization
behaviour for the SFF has been observed before for SYK \cite{Cotler:2016fpe}.

\subsection{Thouless scale and onset of the RMT ramp}
\label{bottleneck}

The spectral form factor also allows us to determine the scale at
which the spectrum first enters the universal random-matrix regime.
For the full physical GUE ensemble, following the criterion of
Ref.~\cite{Gharibyan:2018jrp}, we define the Thouless time
$t_{\rm Th}$ as the earliest robust time beyond which the connected
spectral form factor agrees with the fitted linear RMT ramp within a prescribed
fractional tolerance.  In terms of the unfolded SFF time
\be
 \tau :=\frac{t}{t_H},
 \ee
 where
 \be
 t_H=\frac{2 \pi}{\Delta}
 \ee
 is the Heisenberg time associated with the mean level spacing
 $\Delta$.  It sets the time scale at which individual energy levels
 become resolved and the random-matrix ramp approaches the plateau.
 For this Thouless analysis, we use 4000 independent full-ensemble realizations at each
 $N$ and evaluate the connected SFF on an exact uniform grid with spacing
 $\D \t = 10^{-4}$.  We fit the ramp over 
the interval $0.10\leq\tau\leq0.80$. 
To avoid mistaking a transient crossing for the onset of the RMT ramp,
we define $\tau_{\rm Th}$ as the earliest time after which the connected
spectral form factor remains within a prescribed fractional tolerance
of the fitted ramp $a\t$,
\be
\left|
\frac{K_c(\tau)-a\tau}{a\tau}
\right|<\epsilon.
\ee
For the results quoted below we take a $10\%$ fractional-deviation criterion
$\epsilon=0.10$. 
The resulting values are
\be
\begin{array}{c|cccccccc}
 N
 &12&15&20&25&30&45&50&60\\ \hline
 \tau_{\rm Th}
 &0.0673&0.0559&0.0423&0.0345&0.0276&0.0186&0.0168&0.0133
\end{array}.
\label{tab:thouless-values}
\ee
The product $N\tau_{\rm Th}$ is nearly independent of $N$.  Averaging
over the values of $N$ studied here, we findthe sample mean and
the sample standard deviation:
\be
\overline{N\tau_{\rm Th}} = 0.8322, \qquad {\rm std}(N\tau_{\rm Th})
=0.0208.
\ee
A one-parameter fit gives
\be
\tau_{\rm Th}
 = \frac{0.8283}{N}, \qquad {\rm RMSE} = 8.97 \times 10^{-4}.
\label{eq:thouless-scaling}
\ee
If the exponent is allowed to float, we find 
\be\label{eq:freepower}
 \tau_{\rm Th}=0.7645 N^{-0.9717},  \qquad {\rm RMSE} = 7.71\times10^{-4}.
 \ee
Because the onset definition itself carries a threshold systematic,
the small reduction in residual should not be interpreted as evidence
against $p=1$.  The free exponents for the $5\%$, $20\%$, and $30\%$
criteria are $0.971$, $0.971$, and $0.969$, respectively.  The stable
conclusion is therefore
\be
\label{eq:geomscaling}
\frac{t_{\rm Th}}{t_H}
 = O(N^{-1}).
\ee
For comparison, the alternatives $c/(N\log N)$ and $c/\sqrt N$
give RMSE values $5.45\times10^{-3}$ and $8.54\times10^{-3}$, respectively,
which is considerably worse than the $1/N$ law over the measured range.

It is useful to express this result directly in spectral terms.
The Thouless conductance, the
number of mean level spacings contained within the Thouless energy 
window $E_{\rm Th} := 1/t_{\rm Th}$:
\be 
\label{thouless-cond}
g_{\rm Th}:=\frac{E_{\rm Th}}{\Delta} =\frac{t_H}{2 \pi t_{\rm Th}}
 =O(N).
 \ee
 It means that the Thouless energy extends over an
$O(N)$-level interval. Since each one-parton band itself contains
$O(N)$ levels, the RMT correlations are therefore not confined to
nearest-neighbor spacings but extend over a finite fraction of the
entire fluctuation-induced one-parton spectrum.
The result \eq{eq:geomscaling} further implies that the ramp begins on
the microscopic time scale set by the inverse one-parton bandwidth,
rather than on a new time scale that grows with $N$. Universal
spectral rigidity is therefore established already at this microscopic
scale, with no preceding parametrically slow bottleneck.

Summarizing, the different diagnostics give a consistent picture.  A
typical conditional Hamiltonian exhibits Dyson level repulsion, with
the local statistic $\bar r$ becoming self-averaging as $N$ increases.
The late-time SFF, by contrast, is not self-averaging: its
realization-to-realization fluctuations remain of the same order as
its mean.  Ensemble averaging nevertheless reveals a clean RMT ramp
and hence provides a robust determination of the universal long-range
spectral correlations and their Thouless scale.

\subsection{Pandey--Mehta GOE-to-GUE crossover}
\label{sec:crossover}

That the antiunitary symmetry distinguishes the GUE from the
GOE/GSE classes is already nontrivial.  An even more stringent
question, however, is whether the model exhibits universal spectral
signatures of quantum chaos throughout the continuous breaking of
this symmetry, rather than only at the two endpoints.  In random-matrix
theory this question was answered by Pandey--Mehta, who showed that
gradual breaking of time-reversal symmetry generates a universal
one-parameter family of spectral correlations interpolating between
the orthogonal and unitary ensembles.  The universality therefore
extends to the entire crossover profile, not merely to its GOE and GUE
limits.
The Pandey--Mehta crossover \cite{Pandey:1982br,Mehta:1983ns}
is described by the one-parameter ensemble
\be \label{PM}
H_{\rm PM}(x)
=
S+i\,\frac{x}{\sqrt N}\,A ,
\ee
where $S$ is a real symmetric Gaussian matrix and $A$ is an
independent real antisymmetric Gaussian matrix, with their independent
off-diagonal elements normalized to have the same variance.  At
$x=0$ the Hamiltonian is time-reversal invariant and has GOE
statistics, while increasing $x$ breaks time reversal and drives the
local correlations continuously toward the GUE limit.  The factor
$1/\sqrt N$ is essential: a symmetry-breaking perturbation whose
individual matrix elements vanish with $N$ can nevertheless compete
with the shrinking mean level spacing and produce an $O(1)$ change
in the local spectral statistics.  The finite variable $x$ therefore
resolves the nontrivial GOE-to-GUE crossover in the large-$N$ limit.

The universality of the crossover means that, after unfolding and
expressing the symmetry-breaking perturbation in terms of the
appropriate scaled crossover parameter $x$, the spectral correlation
functions become independent of microscopic details and depend only on
$x$ and on the two Dyson symmetry classes being connected.  This
universal crossover is established within random-matrix theory.

For the microscopic horizon model, the nontrivial question is whether
its strongly structured ensemble belongs to the same
orthogonal-to-unitary crossover universality class.  The endpoint
symmetry classes alone do not guarantee this: additional microscopic
structure could in principle remain relevant throughout the
interpolation and modify the crossover profile.  Comparison with the
Pandey--Mehta crossover therefore provides a stringent test of whether
the structured horizon covariance becomes irrelevant to the fine
spectral correlations once the appropriate crossover scale is
identified.

To analyze this, let us
decompose the Hamiltonian into its time-reversal-even and
time-reversal-odd tensor sectors,
\be
 \mathcal H_s=\mathcal H_{s,e}+\mathcal H_{s,o},
\ee
and introduce
\be
 \mathcal H_s(\eta)
 =
 \mathcal H_{s,e}+\eta\mathcal H_{s,o}.
 \label{eq:crossoverH}
\ee
For even $N$, $\eta=0$ preserves time reversal and gives the GOE
ensemble, while increasing $\eta$ breaks time reversal and drives the
spectrum toward GUE.  We show now that this
provides a microscopic realization of the
Pandey--Mehta orthogonal-to-unitary crossover.

In order to identify the appropriate crossover coordinate $x$, we need
to determine the microscopic strength of the time-reversal-breaking
perturbation relative to the time-reversal-preserving part of the
Hamiltonian.  This information is fixed by the covariance of the
even- and odd-$K$ sectors. Define, for each parton band $s=u,d$,
their total variances carried by
the time-reversal-even and time-reversal-odd sectors,
\be
 \cV_{e,s}
 :=
 \left\langle\Tr\mathcal H_{s,e}^2\right\rangle_a,
 \qquad
 \cV_{o,s}
 :=
 \left\langle\Tr\mathcal H_{s,o}^2\right\rangle_a .
\ee
Using  the orthonormality of the tensor basis, 
\be
 \Tr\!\left[
 \Mten_{K,A}^{(k_s)\dagger}\Mten_{K',A'}^{(k_s)}
 \right]
 =
 \delta_{KK'}\delta_{AA'},
\ee
these are
\be
 \cV_{e,s}
 =
 \sum_{K\ {\rm even}}(2K+1)w_{K,s}^2,
 \qquad
 \cV_{o,s}
 =
 \sum_{K\ {\rm odd}}(2K+1)w_{K,s}^2 .
\ee
The corresponding band-resolved symmetry-breaking strength is
\be
 \rho_s^2:=\frac{\cV_{o,s}}{\cV_{e,s}}.
\ee
Since the crossover statistic reported below is obtained by combining
the separately evaluated $u$- and $d$-band results, we introduce a
single pooled  symmetry-breaking strength
\be\label{rhos-def}
 \rho^2
 :=
 \frac{\cV_{o,u}+\cV_{o,d}}
      {\cV_{e,u}+\cV_{e,d}} .
\ee
This pooling is used only to define the common crossover coordinate.
The two spectra themselves remain distinct throughout the spectral
analysis.
Note that $\rho_s$ and $\rho$ are fixed entirely by
the microscopic vacuum mode spectrum and projection coefficients.

The bare deformation parameter $\eta$ is not itself the natural
finite-size crossover variable, because the onset of symmetry breaking
also depends on the matrix dimension.  The relevant criterion is that
the typical time-reversal-breaking matrix element becomes comparable to
the local mean level spacing,
\be
 \eta\left|(\mathcal H_{s,o})_{mn}\right|_{\rm rms}
 \sim \Delta .
\ee
For a matrix of dimension $d$ and a spectral width $W$,
\be
 \Delta\sim\frac{W}{d}.
\ee
On  the other hand, from the definition \eq{rhos-def}
and that $\langle{\rm Tr}\, \cH_{s,e}^2\rangle\sim d \, W^2$, one has
$\langle{\rm Tr}\,\cH_{s,o}^2\rangle\sim\rho^2 d\,W^2$.
If the odd-sector strength is distributed over $O(d^2)$ matrix elements
in the eigenbasis of $\cH_{s,e}$, a typical symmetry-breaking
matrix element therefore scales as
\be
 |(\cH_{s,o})_{mn}|_{\rm rms}
 \sim \frac{\rho W}{\sqrt d}.
\ee
Consequently, the ratio of the symmetry-breaking mixing energy to the mean level
spacing is 
\be
 \frac{
 \eta\left|(\mathcal H_{s,o})_{mn}\right|_{\rm rms}}
 {\Delta}
 \sim \eta\rho\sqrt d .
\ee
The crossover occurs when this ratio becomes of the order of unity. 
Since the two parton bands have dimensions
$d_u=N+1$ and $d_d=N-1$, the natural finite-size crossover coordinate is
therefore
\be
 x:=\eta\rho\sqrt N .
 \label{eq:xscale}
\ee
Thus $x\ll1$ corresponds to the approximately time-reversal-invariant
GOE regime, $x=O(1)$ to the crossover region, and $x\gg1$ to the
time-reversal-broken GUE regime.

For the finite-$N$ branch-resolved horizon ensemble, the microscopic
covariance gives
\be
 \rho_{20}=0.9570,\qquad
 \rho_{30}=0.9721,\qquad
 \rho_{60}=0.9866.
 \label{rho-table}
\ee
We evaluate the crossover on the common scaled grid
\be
 x=0,0.05,\ldots,2.50,
\ee
using $4000$ microscopic realizations for $N=20$ and $30$ and $1000$
for $N=60$.  For each $N$, the corresponding bare deformation is
\be
 \eta=\frac{x}{\rho\sqrt N},
\ee
so that the three ranks are compared at the same physical crossover
strength rather than at the same value of $\eta$.

\begin{figure}[ht]
\centering
\includegraphics[width=0.96\textwidth]
                {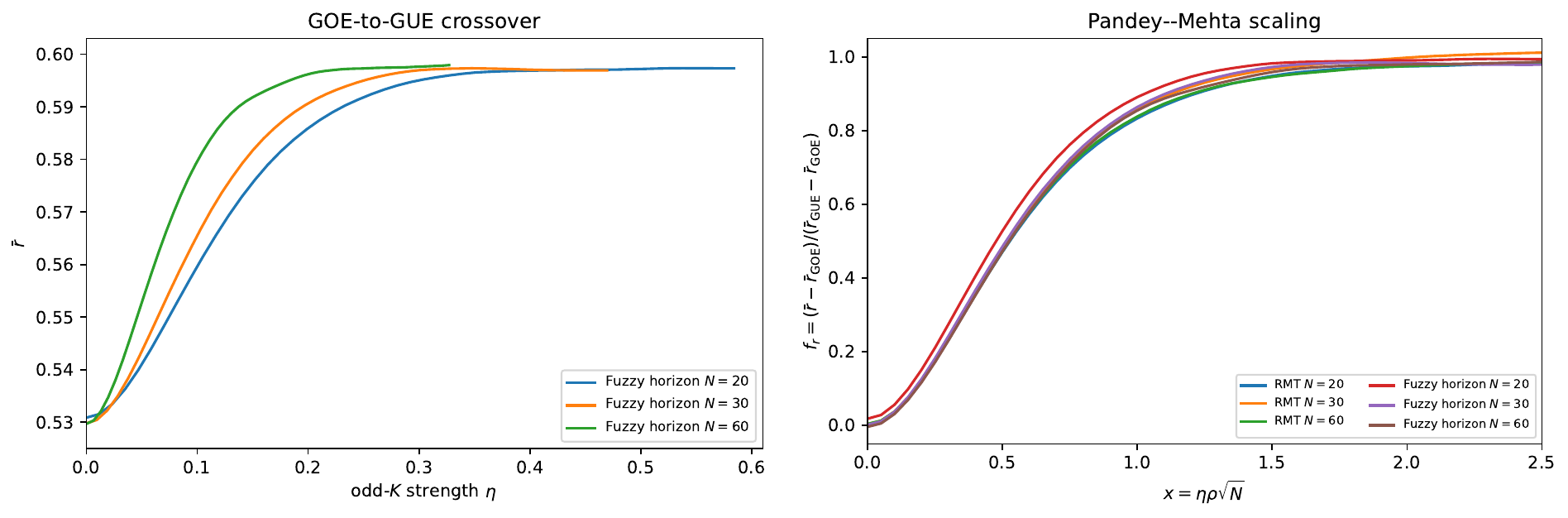}
\caption{GOE-to-GUE crossover of the finite-$N$ branch-resolved
microscopic horizon Hamiltonian.
Left: pooled adjacent-gap ratio $\bar r$ as the odd-$K$ sector is
turned on by the bare deformation $\eta$.
The different terminal values of $\eta$ correspond to the common
scaled range $0\leq x\leq2.5$.
Right: normalized crossover fraction $f_r$ versus the microscopic
coordinate $x=\eta\rho\sqrt N$, together with the corresponding
matched finite-dimensional Pandey--Mehta curves.
The covariance ratio $\rho$ is determined microscopically, without any
additional normalization introduced by hand.}
\label{fig:crossover}
\end{figure}

Figure~\ref{fig:crossover}(a) shows the resulting adjacent-gap statistic.
The bare-$\eta$ curves interpolate monotonically from the GOE to the GUE regime.
To compare the different ranks quantitatively, we define
\be
 f_r(x)
 :=
 \frac{
 \bar r(x)-\bar r_{\rm GOE}^{(N)}
 }{
 \bar r_{\rm GUE}^{(N)}-\bar r_{\rm GOE}^{(N)}
 },
 \label{eq:fr}
\ee
where $\bar r_{\rm GOE}^{(N)}$ and $\bar r_{\rm GUE}^{(N)}$ are the
corresponding matched finite-dimensional reference values.  With this
normalization, $f_r=0$ and $f_r=1$ represent the finite-dimensional GOE
and GUE values, respectively.
As shown in Figure~\ref{fig:crossover}(b),
the microscopic curves for the three ranks collapse closely when
plotted against $x$.  Defining $x_{1/2}$ by
\be
 f_r(x_{1/2})=\frac12 ,
\ee
we obtain
\be
 x_{1/2}
 =
 0.477\quad (N=20),\qquad
 0.513\quad (N=30),\qquad
 0.525\quad (N=60).
 \label{x12}
\ee
The extended range also shows explicitly that the crossover reaches the
unitary regime: at $x=2.5$ the microscopic values of $f_r$ are
$0.994$, $0.979$, and $0.987$ for $N=20$, $30$, and $60$,
respectively.

For comparison, we construct matched finite-dimensional
Pandey--Mehta ensembles \eq{PM} with the same two matrix dimensions,
$d_u=N+1$ and $d_d=N-1$.   Since the adjacent-gap ratio is
invariant under a common rescaling of the spectrum, the overall
variance is immaterial.  We therefore normalize the independent off-diagonal
elements of $S$ and $A$ to have the same variance, thereby fixing the
normalization of $x$.  At every value of $x$,
$\bar r_{\rm PM}(x)$ is evaluated separately for the
two dimensions $d_u$ and $d_d$, and pooled with the same gap-count
weights used for the microscopic Hamiltonian.
The Pandey--Mehta calculation uses the same numbers of
realizations as the corresponding horizon calculation.
Using the same finite-dimensional normalization
\eqref{eq:fr}, the Pandey--Mehta crossover points are
\be
 x_{1/2}^{\rm PM}
 =
 0.527\quad (N=20),\qquad
 0.518\quad (N=30),\qquad
 0.525\quad (N=60).
 \label{x12PM}
\ee
The discrepancy between the microscopic crossover values \eqref{x12}
and the reference crossover points \eqref{x12PM} decreases from
about $0.050$ at $N=20$ to $0.005$ at $N=30$, and is indistinguishable
at the quoted precision at $N=60$. Moreover, the comparison is not
restricted to the midpoint: the complete microscopic and
Pandey--Mehta curves track one another throughout the crossover and
both approach the finite-dimensional GUE value at large $x$.
This agreement is highly nontrivial.  The GOE and GUE endpoint
symmetry classes alone do not determine the microscopic interpolation
between them.  Indeed, Kumar and Pandey showed that microscopically
distinct orthogonal-to-unitary crossover ensembles, including the
Gaussian, Laguerre, and Jacobi ensembles, have different
finite-dimensional correlation functions. Only after unfolding and
an appropriate rescaling of the symmetry-breaking parameter do their
large-$N$ spectral correlations collapse onto the same universal
Pandey--Mehta  crossover \cite{KumarPandey:2009}.  The horizon
ensemble is considerably more structured than these random-matrix
examples: its operator basis is fixed by the fuzzy-sphere geometry and
its covariance is strongly rank dependent.
It is therefore
nontrivial that, once the microscopic time-reversal-breaking strength
is expressed in terms of the scaled variable
$x=\eta\rho\sqrt N$, with $\rho$ determined independently from the
geometric covariance, the horizon ensemble follows the same
Pandey--Mehta crossover profile without spectral unfolding or any
additional $N$-dependent fitting parameter.  At the level of the
adjacent-gap statistic, this demonstrates that the detailed geometric
structure of the horizon Hamiltonian does not generate a distinct
orthogonal-to-unitary crossover: its microscopic spectral evolution
flows onto the universal Pandey--Mehta trajectory.

\subsection{Rank-resolved variance and operator support}
\label{operator_support}

The emergence of random-matrix statistics raises the question of how
broadly the quantum horizon fluctuations act on the one-parton Hilbert
space.  For a given fermion band $s$, the covariance weight of one
component of the rank-$K$ tensor multiplet is given by \eq{wKs},
where the two physical branches contribute independently.  Since a
rank-$K$ multiplet contains $2K+1$ Hilbert--Schmidt-orthonormal real
components, the total Hamiltonian variance carried by that rank is
\be
 \cV_{K,s}=(2K+1)w_{K,s}^{2}.
 \label{eq:rank-variance}
\ee
All quantities here are evaluated with the complete finite-$N$
quantum-corrected curvatures used in the spectral calculation.
To characterize how the fluctuation strength is distributed
over tensor ranks, 
we introduce the cumulative band variance defined by
\be
 F_s(K):=\sum_{K'\leq K}f_{K',s},
 \qquad
 f_{K,s}
 :=
 \frac{\cV_{K,s}}
 {\displaystyle\sum_{K'}\cV_{K',s}} .
 \ee
 
\begin{figure}[t]
 \centering
 \includegraphics[width=0.82\linewidth]{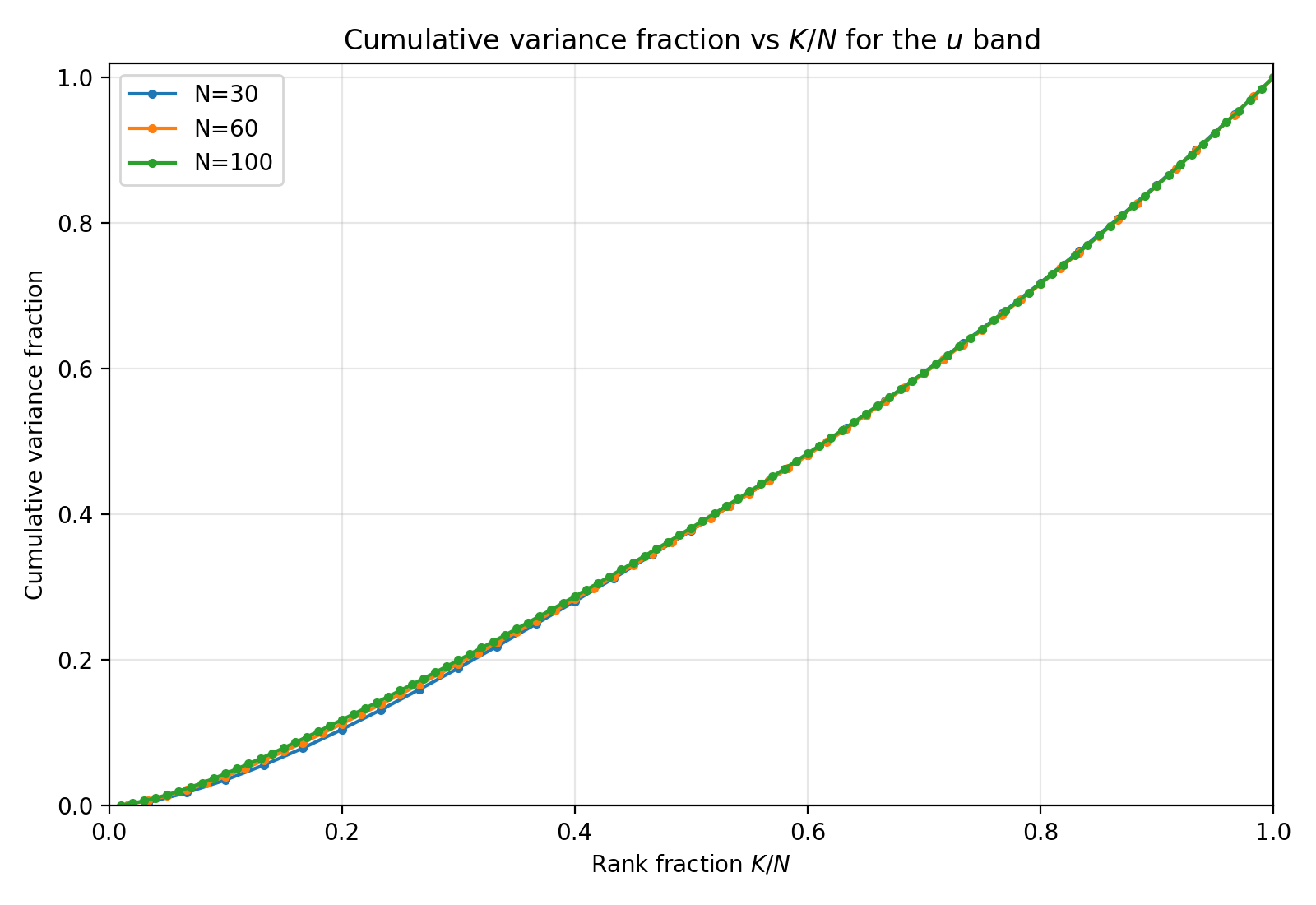}
 \caption{Cumulative fraction $F_u(K)$ of the fluctuation-induced
 $u$-band Hamiltonian variance as a function of $K/N$, computed with
 the complete finite-$N$ quantum-corrected curvatures.  The approximate
 collapse shows that an $O(N)$ range of tensor ranks remains active as
 $N$ increases.}
 \label{fig:rank-cumulative}
\end{figure}

Figure~\ref{fig:rank-cumulative} shows the cumulative $u$-band
variance
as a function of the scaled rank $K/N$.  The curves for different $N$
collapse increasingly well, showing that the fluctuation strength
remains broadly distributed over an $O(N)$ range of tensor ranks
rather than concentrating into a fixed set of long-wavelength modes.
At $N=100$, only $4.4\%$ of the $u$-band variance lies at
$K\leq\sqrt N$, whereas $48.4\%$ already lies in the ultraviolet
window $K\geq N^{0.9}$. The cumulative variance reaches
$50 \%$ only at $K=62$, corresponding to $K/N\simeq0.62$, 
while the largest single-rank contribution, from $K=100$, is only
$1.55\%$.  The even- and odd-rank sectors carry $50.4\%$ and $49.6\%$
of the variance, respectively.  Thus neither a small set of soft modes
nor one symmetry sector dominates the fluctuation-induced Hamiltonian.

The broad distribution over tensor rank can also be understood
analytically at large $N$.  Consider the scaling limit
\be
\k:=\frac{K}{N}
\ee
with $0<\k<1$ held fixed.  For the $u$-parton band, the two physical
branches contributing to a given tensor rank $K$ satisfy
\be \label{RRR}
\left|\mathcal R^{(u)}_{K-1,K}\right|^2
\longrightarrow \frac{(1+\k)^2}{2},
\qquad
\left|\mathcal R^{(u)}_{K+1,K}\right|^2
\longrightarrow \frac{(1-\k)^2}{2},
\ee
while their effective curvatures have the common leading behavior
\be
\Lambda^{\rm eff}_{K\pm1,K}\longrightarrow \k^2N^2 .
\ee
For the $d$-parton, the expression \eq{RRR} is swapped with $\k \to -\k$.
It follows from \eqref{wKs} that
\be
w_{K,s}^2
\longrightarrow
\frac{1+\k^2}{\k N}, \quad s=u, d,
\ee
and hence the total variance carried by rank $K$ becomes
\be
\mathcal \cV_{K,s}=(2K+1)w_{K,s}^2
\longrightarrow 2(1+\k^2).
\label{eq:Vrank-largeN}
\ee
Thus no fixed range of low tensor ranks carries a finite fraction of
the variance as $N\to\infty$. Instead, an $O(1)$ fraction remains
distributed over ranks $K=O(N)$.  Treating $\k=K/N$ as continuous, the
corresponding normalized cumulative distribution is
\be
F(\k)
=\frac{\displaystyle\int_0^\k dy\,2(1+y^2)}
{\displaystyle\int_0^1dy\,2(1+y^2)}
=\frac{3\k+\k^3}{4}.
\label{eq:Frank-largeN}
\ee
This large-$N$ result is the same for the two parton bands and explains
the approximate finite-$N$ collapse seen in
Figure \ref{fig:rank-cumulative}.
For example,
\be
1-F\left(\frac12\right)\simeq0.594,
\ee
so asymptotically about $59.4\%$ of the fluctuation-induced Hamiltonian
variance is carried by tensor ranks $K\ge N/2$.

A useful quantitative measure of the breadth of the tensor-rank support is
the participation ratio of the rank-resolved variance,
\be
N_{\rm rank}^{\rm eff}
:=
\frac{\left(\sum_K {\cal V}_{K,s}\right)^2}
{\sum_K {\cal V}_{K,s}^2}.
\ee
This quantity is normalization independent and is bounded above by the
number of available ranks, with the maximum attained for a
rank-independent variance profile.  Using the large-$N$ result
${\cal V}_{K,s}\to2(1+\kappa^2)$, with $\kappa=K/N$, we obtain
\bea
\sum_K {\cal V}_{K,s}
&\simeq&
N\int_0^1 d\kappa\,2(1+\kappa^2)
=\frac{8}{3}N,
\\
\sum_K {\cal V}_{K,s}^2
&\simeq&
N\int_0^1 d\kappa\,4(1+\kappa^2)^2
=\frac{112}{15}N,
\eea
and hence
\be
N_{\rm rank}^{\rm eff}
\simeq
\frac{20}{21}N.
\ee
For the $u$ band, the finite-$N$ sums approach this limit from below,
giving
$N_{\rm rank}^{\rm eff}/N=0.880$, $0.923$, $0.944$ at
$N=12,\,30,\,100$, respectively.  The near-maximal value arises
because ${\cal V}_{K,s}$ remains of order unity over the entire rank
range, varying only between $2$ and $4$ in the large-$N$ limit.
The small deficit from a completely flat profile is due to the
$\kappa^2$ dependence, which originates from the two branch factors
in \eqref{RRR}; for the $d$ band the two factors are interchanged.

For comparison, in an invariant GUE,   after removal of the identity
component, the variance is uniform per
Hilbert--Schmidt-normalized tensor component.  Since rank $K$ contains $2K+1$
components, its total rank-resolved variance is therefore
\be
{\cal V}^{\rm GUE}_K=(2K+1)w^2
\propto \kappa N .
\ee
The corresponding cumulative distribution is
$F_{\rm GUE}(\kappa)=\kappa^2$, so that one half of the variance lies
above $\kappa=1/\sqrt{2}\simeq0.71$, compared with
$\kappa\simeq 0.60$ for the horizon ensemble.  The associated rank
participation ratio is
\be
N_{\rm rank}^{\rm eff}\big|_{\rm GUE}
=
\frac{\left(\int_0^1 d\kappa\,2\kappa\right)^2}
{\int_0^1 d\kappa\,4\kappa^2}\,N
=
\frac34 N .
\ee
Thus the horizon covariance produces an unusually flat distribution
of the total Hamiltonian variance across tensor ranks: a near-maximal
$O(N)$ number of ranks contributes appreciably.  This should not be
interpreted as the horizon ensemble being more random than GUE.
Rather, its microscopic rank-dependent covariance compensates much of
the channel multiplicity that makes the GUE rank profile grow linearly
with $K$.  The comparison shows that, although an $O(N)$ range of
tensor ranks is also active in GUE, the near-uniform distribution of
variance across ranks in the horizon ensemble is a nontrivial
consequence of its microscopic covariance. Together with
$N_{\rm rank}^{\rm eff}\simeq(20/21)N$, this establishes that the
fluctuation-induced Hamiltonian has macroscopically broad tensor-rank
support in the large-$N$ limit.  The implications of this broad
operator support for nonlocality and fast scrambling are discussed in
Section \ref{fast_scrambling}.

We remark that the emergence of random-matrix universality can be
understood from a specific property of the fuzzy-sphere fluctuation
spectrum: the high-$K$ modes remain sufficiently soft that the induced
one-parton Hamiltonian has support over an $O(N)$ range of tensor
ranks and is therefore effectively full rather than parametrically
banded.
Had the fluctuation strength instead been concentrated at low tensor
rank, the induced Hamiltonian would have become parametrically banded,
a regime in which a crossover away from Wigner--Dyson and toward
Poisson statistics can occur \cite{Mirlin:1996zz}.

\section{Physical Nature of Geometry-Induced Chaos
}
\label{sec:physical}

The numerical results of the preceding section establish that the
quantum-geometric-fluctuation-induced ensemble exhibits the spectral
correlations of the appropriate Dyson classes, together with the
Pandey--Mehta crossover when time-reversal symmetry is continuously
broken.  We now turn to the physical nature of this chaos.  We first
clarify how universal random-matrix correlations emerge from a highly
structured microscopic ensemble whose probability law is generated by
the quantum state of the horizon itself.  We then discuss how the
conditional one-parton Hamiltonian acquires an operational
interpretation through decoherence and, more generally, how
time-dependent horizon fluctuations may promote the instantaneous
spectral chaos found here into dynamical response.  Finally, we
distinguish this geometry-induced single-parton chaos from the
additional many-parton chaos that may arise through interactions, and
discuss the corresponding Thouless, Lyapunov, and scrambling scales.

\subsection{From microscopic dynamics to universality}

An important feature of the present construction is that the microscopic
ensemble is not itself an invariant random-matrix ensemble.  The
Hamiltonian has the structured form
\be
 \cH_s(\bm\xi)
 =
 \sum_{K,A}Y^{(s)}_{K,A}(\bm\xi)\,
 \Mten^{(k_s)}_{K,A},
\ee
where the tensor matrices $\Mten^{(k_s)}_{K,A}$ are fixed by the
fuzzy-sphere geometry and the random amplitudes have the nontrivial
rank-dependent covariance
\be
 \left\langle
 Y^{(s)}_{K,A}Y^{(s)\,*}_{K',A'}
 \right\rangle
 =
 w_{K,s}^2\,\delta_{KK'}\delta_{AA'} .
\ee
Thus both the operator basis and the covariance retain detailed
microscopic information about the horizon geometry.

The numerical results show, however, that this microscopic structure
does not survive in the universal spectral correlations.  Once the
antiunitary symmetry is specified, the local level statistics and the
connected spectral correlations follow the corresponding finite-dimensional
GOE, GUE, or GSE benchmarks.  The microscopic covariance is
nevertheless not irrelevant:  it determines how the physical
Hamiltonian enters the universal regime.  This is particularly clear in
the GOE-to-GUE crossover, where the covariance fixes the relative
time-reversal-breaking strength $\rho$ and hence the mapping
\be
 \eta
 \longmapsto
 x=\eta\rho\sqrt N .
\ee
The map 
from the microscopic deformation $\eta$ to $x$ is model dependent,
whereas the dependence of the spectral correlations on $x$ becomes universal.

The horizon theory thus provides a microscopic realization of
random-matrix universality. The fuzzy sphere geometry and its
quantum fluctuations determine the
structured ensemble, while the antiunitary
symmetry selects the appropriate Dyson class.  The microscopic
covariance fixes the strength of the time-reversal-breaking sector and
therefore the mapping to the universal Pandey--Mehta crossover
coordinate.  Once these microscopic inputs are specified, the detailed
fuzzy-sphere structure becomes invisible to the spectral
diagnostics tested here, once the symmetry class and crossover
coordinate are fixed.

It is useful in this regard to contrast the present construction with
SYK.  In SYK, a realization of the couplings $J_A$ is drawn from a
prescribed quenched probability distribution and defines a many-body
Hamiltonian $H_{\rm SYK}(J)$.  Here, by contrast, the variables
$q_{LK,A}$ are dynamical quantum coordinates of the microscopic
horizon geometry, and their probability distribution follows from the
horizon wavefunction itself.  Schematically,
\be
{\rm SYK}:\qquad P(J)\longrightarrow H_{\rm SYK}(J),
\qquad\qquad
{\rm horizon}:\qquad
|\Psi_0(q)|^2\longrightarrow H_{\rm p}(q).
\ee
Thus randomness is an input in the standard definition of SYK,
whereas in the present construction the ensemble is generated by the
quantum state of spacetime.  Moreover, the resulting horizon ensemble
is not invariant in matrix space: its covariance retains the
rank-dependent tensor structure dictated by the fuzzy geometry.
The emergence of Dyson universality is therefore not a consequence of
postulating a Gaussian random Hamiltonian, but of the loss of
microscopic information in sufficiently fine spectral correlations.

\subsection{Single-parton chaos and the multi-parton extension}

Black-hole quantum chaos is usually discussed as a many-body
phenomenon with standard features such as scrambling, operator growth,
out-of-time-order correlators, and the emergence of random-matrix
statistics in the exponentially large many-body Hilbert space.  The
chaos found here has a different microscopic origin.  Even before
interactions among many horizon partons are considered, the
Hamiltonian of a single parton conditioned on the quantum state of the
horizon already exhibits the spectral correlations of the appropriate
Dyson ensemble.  In this sense the model contains a genuinely
single-parton form of quantum chaos induced directly by fluctuations
of the microscopic spacetime geometry.

The situation is reminiscent, in spirit, of Wigner's use of random
matrices to describe complicated nuclear spectra.  An observed degree
of freedom is sensitive to a large number of unresolved microscopic
degrees of freedom, whose detailed dynamics is effectively replaced
by universal spectral statistics.  Here, however, the unresolved
degrees of freedom are not internal matter constituents but the
quantum-geometric modes of the horizon itself.  A horizon parton
therefore experiences a conditional Hamiltonian determined by the
unobserved ``cloud'' of noncommuting horizon fluctuations.  The
important difference is that, whereas random-matrix theory provides an
effective statistical description of the unresolved microscopic dynamics
in the nuclear problem, here the statistical ensemble is derived directly
from the quantum theory, with its probability distribution determined
explicitly by the quantum ground state of the horizon geometry.

This single-parton chaos does not exhaust the possible chaotic dynamics
of the horizon.  Once several partons are present, the same microscopic
degrees of freedom also mediate interactions among them.  Schematically,
the many-parton Hamiltonian takes the form
\be \label{H-int}
H_{\rm MB}=
\sum_{ij} h_{ij}({\bf q})c_i^\dagger c_j + H_{\rm int},
  \ee
  where $h({\bf q})$ is the geometry-induced chaotic one-parton
  Hamiltonian studied above, while $H_{\rm int}$ contains genuine
  many-parton interactions.  Integrating out the bosonic horizon
  fluctuations generates quartic fermion interactions reminiscent of
  those in SYK.  The analogy is nevertheless only structural: here the
  effective four-fermion couplings are not independent random variables,
  but have the factorized form
  \be \label{J4}
  J_{ij;kl}(\o) =
\sum_A G_A(\o) (T_A)_{ij}(T_A)_{kl},
\ee
fixed by the spectrum and tensor structure of the quantum horizon
geometry. Remarkably, the same operators $T_A$ that produce the
geometry-induced single-parton mixing also mediate scattering in the
many-parton Hilbert space.

We emphasize that the two forms of chaos are distinct but related. The
geometry-induced one-parton RMT structure is embedded in the
many-body Hamiltonian, interactions allow the partons to explore
the much larger Fock space and may generate additional many-body chaos
associated with thermalization, operator growth, and scrambling.
The corresponding diagnostics are also distinct. Whereas the
Thouless time
$t_{\rm Th}^{\rm geom}$, due to the horizon geometric fluctuations,
measures the onset of universal spectral
correlations in the conditional one-parton spectrum,  the
scrambling time $t_{\rm scr}$ characterizes the real-time information
spreading of quantum information.
There is no general relation between these scales. The
interacting system may also possess its own many-body Thouless time
$t_{\rm Th}^{\rm MB}$. The exhibition of the many-body RMT universality,
ETH, and fast scrambling in the horizon
parton model is a separate dynamical question for
future work.

\subsection{Decoherence and dynamics beyond the conditional spectrum}

The conditional ensemble discussed above originates from a quantum
state of the horizon geometry rather than from classical disorder.
Thus, in a closed system, the different configurations $\bm q$ appearing
in the Born distribution should not in general be regarded as an
incoherent statistical ensemble.  A physical black-hole horizon,
however, is not isolated.  If the geometric degrees of freedom become
entangled with unobserved environmental degrees of freedom, tracing
over the latter suppresses the coherence between distinct geometric
configurations.  Schematically, for an interaction
\be
 H_{\rm int}
 =
 \sum_\alpha q_\alpha F_\alpha ,
\ee
the reduced density matrix evolves as
\be
 \rho_{\rm geom}(\bm q, \bm q';t)
 =
 \rho_{\rm geom}(\bm q,\bm q';0)
 \langle E_{\bm q'}(t)|E_{\bm q}(t)\rangle ,
\ee
where $|E_{\bm q}(t)\rangle$ denotes the environmental state correlated with
the configuration $\bm q$.  When this overlap becomes small for
$\bm q\neq \bm q'$, the reduced geometric state becomes approximately diagonal,
\be
 \rho_{\rm geom}
 \simeq
 \int d \bm q\,
 P(\bm q)\,|\bm q\rangle\langle \bm q|,
 \qquad
 P(\bm q)\simeq |\Psi_0(\bm q)|^2 ,
\ee
up to corrections from backreaction and environmental renormalization.
In this regime, the Born distribution acquires a direct operational
interpretation as an effective ensemble of geometric backgrounds.

The matrix model contains a natural microscopic candidate for the
environment responsible for such decoherence.  In the two-block
construction, off-diagonal link modes connect the horizon block to an
exterior block and condense near the horizon.  The same link sector
that dynamically locks the horizon and exterior variables also
entangles them \cite{Chu:2026vzj,Chu:2026qom}.
Expanding about the condensed configuration
generically produces interactions of the form
\be
 H_{\rm int}^{\rm link}
 =
 \sum_\alpha q_\alpha F_\alpha(Z,E)+\cdots ,
\ee
where $E$ collectively denotes the exterior
environmental degrees of freedom, and 
$Z_a$ denotes the off-diagonal link fields connecting the horizon
and exterior matrix blocks. As a result,
different horizon geometries become correlated with different
states of the link and exterior sectors.  Tracing over these degrees of
freedom therefore provides a microscopic route to decoherence in the
geometric configuration basis. It is interesting to
determine the corresponding decoherence rate.
This will be left for future consideration.
We emphasize that decoherence is not an assumption required
for the random-matrix analysis above, but rather a possible physical
mechanism by which the quantum-geometric Born distribution can acquire
the interpretation of an effective statistical ensemble of conditional
Hamiltonians.

If the decohered geometry is also sufficiently slow on the time scale
of the parton observable, the conditional Hamiltonian is approximately
quenched.  More generally, however, the effective geometric
configuration may evolve appreciably, in which case the partons
experience a time-dependent Hamiltonian
\be
 H_{\rm p}(t)=H_{\rm p}[\bm q(t)] .
\ee
The spectral results obtained here then characterize the instantaneous
Hamiltonian along this evolution.  Its time dependence opens the
possibility of additional dynamical phenomena: energy levels may
wander, avoided crossings may be traversed, and the fluctuating
geometry may induce transitions and dephasing.  The appropriate
description then becomes one of parametric random-matrix theory or
quantum noise acting on an instantaneously chaotic spectrum.
These dynamical effects are not needed for the static spectral
universality established in this work.  They instead point toward the
next step in relating the one-particle spectral chaos found here to
dynamical phenomena such as relaxation, dissipation, and transport, and
ultimately to stronger notions of quantum chaos in the full interacting
horizon system.  Addressing these questions requires a dynamical
treatment of the geometric, link, and parton sectors together,
which lies beyond the scope of the present study.

\subsection{Nonlocality of the horizon Hamiltonian
  and fast scrambling}
\label{fast_scrambling}

As shown in Section~\ref{operator_support},
the combination of an $O(N)$ rank range and the near-maximal
participation ratio $N_{\rm rank}^{\rm eff}\simeq(20/21)N$ shows
quantitatively that the fluctuation-induced Hamiltonian is supported
on a macroscopically broad set of tensor scales.  This provides a
microscopic perspective on the fast-scrambling conjecture of Sekino
and Susskind \cite{Sekino:2008he}.  According to their conjecture,
black holes are the fastest scramblers in nature, with
$t_{\rm scr}\sim\beta\log S$.  Sekino and Susskind further argued that
such logarithmic scrambling cannot be achieved by systems with local
interactions, for which information propagation is limited by spatial
locality and the scrambling time grows as a power of the system
size~\cite{Sekino:2008he}; see also \cite{Lashkari:2011yi}.  Their
prototype of a fast scrambler was matrix quantum mechanics.  The
present horizon model has this matrix character by construction, while
the rank analysis of Section~\ref{operator_support} shows explicitly
that a macroscopically broad set of tensor channels remains active at
large $N$.

This conclusion is also reflected directly in the matrix structure.
The selection rule $m'=m+Q$, with $|Q|\leq K$, implies that a
rank-$K$ tensor can connect states separated by as much as $K$ in
magnetic quantum number.  Since an $O(1)$ fraction of the variance is
carried by $K=O(N)$ ranks, the induced Hamiltonian contains matrix
elements extending over an $O(N)$ span of the one-parton band rather
than being confined to a narrow band about the diagonal.  This
provides a microscopic interpretation of, and is consistent with, the
absence of a parametrically slow diffusive bottleneck found in the
Thouless analysis, where $t_{\rm Th}/t_H\sim N^{-1}$.  The broad
operator connectivity and the parametrically early onset of
random-matrix universality thus give two complementary indications of
the effectively nonlocal dynamics expected in a fast-scrambling
matrix system.

These structural results do not by themselves establish the
scrambling law, since that requires the interacting many-parton
dynamics and an OTOC or equivalent operator-growth diagnostic.
They do, however, verify directly a central structural ingredient of
the Sekino--Susskind picture in the microscopic horizon model.  The
fluctuation-induced Hamiltonian acts through an $O(N^2)$ set of
noncommuting operator directions on the one-parton Hilbert space and
is consequently neither sparse nor dominated by low-rank channels.
Instead, it acts through a macroscopically broad set of noncommuting
matrix operators, producing highly connected and effectively nonlocal
dynamics on the one-parton Hilbert space.

The matrix horizon model therefore possesses microscopic features of
precisely the kind expected to make matrix quantum mechanics a natural
candidate for fast scrambling.  Whether the full interacting horizon
actually saturates $t_{\rm scr}\sim\beta\log S$, however, remains a
dynamical question for the many-parton theory.

\subsection{Antiunitary symmetry and quantum geometry}
\label{sec:antiunitary-geometry}

The relation between antiunitary symmetry and random-matrix
universality acquires a geometric meaning in quantum
gravity.  In the Stanford--Witten analysis of JT gravity
\cite{Stanford:2019vob}, the antiunitary symmetry of the boundary
quantum system is reflected holographically in the orientation
structure of the bulk gravitational path integral.  In particular,
time-reversal symmetry requires the inclusion of orientation-reversing,
and hence non-orientable, bulk geometries.  The corresponding discrete
symmetry data determine whether the boundary Hamiltonian belongs to
the orthogonal, unitary, or symplectic random-matrix class.

A closely related relation arises microscopically in the present
horizon model.  We start by observing that the  full
matrix horizon theory possesses an antiunitary symmetry
acting simultaneously on the fermionic partons and on the matrix
geometry. Consider a generalized time-reversal transformation  ${\mathsf T}$
defined by
combining the ordinary antiunitary time-reversal operation with the
antipodal inversion of the fuzzy-sphere matrix geometry,
\be
  X_a(t) \longrightarrow
    -\,U_J X_a^{*}(-t) U_J^\dagger , \qquad  U_J = e^{-i\pi J_y}.
    \label{eq:TR-X}
\ee
The additional minus sign is essential: ordinary Wigner time reversal
gives $U_JJ_a^*U_J^\dagger=-J_a$, while the antipodal inversion
supplies a second minus sign. Consequently the round fuzzy-sphere
background $X_a=J_a$ is invariant under the combined transformation
\eqref{eq:TR-X}.  The fermion
transforms under the corresponding combined orbital-spin
antiunitary operation,
\be
\psi(t) \longrightarrow (U_J \otimes i\sigma_2)\psi^*(-t).
\ee
Using
$    (i\sigma_2)\sigma_a^{*}(-i\sigma_2)=-\sigma_a $, 
one sees that the two minus signs cancel in the Yukawa coupling
$\psi^\dagger \sigma^a X_a \psi$.  The bosonic kinetic and potential
terms are likewise invariant.  As a result,
the generalized antiunitary transformation
is a
symmetry of the full boson--fermion matrix model.

To see how this symmetry acts on the quantum geometry, expand the
fluctuation about the fuzzy sphere as
\be
    \delta X_a
    =
    \sum_{L,K,A}
    q_{LK,A}\,B_a^{LK,A}.
    \label{eq:deltaX-mode-expansion}
\ee
Under the antiunitary operation, an orbital rank-$L$ fuzzy spherical
harmonic carries a factor $(-1)^L$, while the explicit minus sign in
\eqref{eq:TR-X} supplies one additional factor.  Since the physical
vector branches satisfy
$    K=L\pm1 $,  therefore
$ (-1)^{L+1}=(-1)^K $.
In the real Hermitian tensor basis used throughout the numerical
analysis, the geometric normal coordinates consequently transform as
\be
 q_{LK,A}(t)  \longrightarrow (-1)^K q_{LK,A}(-t).
    \label{eq:TR-q}
\ee
The even-$K$ and odd-$K$ fluctuations are thus respectively even and
odd under the antiunitary symmetry.

For the spectral problem, however, we condition on a fixed realization
of the quantum geometry and study the corresponding one-parton
Hamiltonian. The reversal of the time argument then plays no direct
role, what remains is the induced action of the antiunitary symmetry on
the geometric configuration,
\be
q_{LK,A}\longrightarrow (-1)^K q_{LK,A}.
\ee
This relation makes an important distinction transparent.  The full
theory is antiunitary invariant, but a generic conditional
Hamiltonian need not be.  A generic quantum-geometric configuration
contains both even- and odd-$K$ components, and the
antiunitary operation then maps $h_s(\mathbf q)$ into the
Hamiltonian $h_s(\mathsf T\mathbf q)$
associated with the distinct geometry. Consequently,
a fixed conditional Hamiltonian is generally not
invariant under this antiunitary operation and hence is not constrained
by it. In the absence of any additional antiunitary symmetry, the
expected Dyson class is unitary, as confirmed by the GUE statistics.
By contrast, if only even-$K$ modes are retained, the geometric
configuration is invariant under the antiunitary transformation, and
the conditional Hamiltonian inherits the corresponding symmetry. For
even $N$ this gives the orthogonal class, GOE, whereas for odd $N$ the
spectrum of distinct Kramers pairs belongs to the GSE symplectic class.
Thus the geometric realization determines whether the
antiunitary symmetry is preserved or broken, while, when it is
preserved, the parity of $N$ distinguishes the GOE and GSE classes.

The resulting picture closely parallels the Stanford--Witten relation
between antiunitary symmetry and quantum geometry.  In JT gravity the
connection is holographic: the antiunitary symmetry of the boundary
theory is encoded in the orientation structure of the bulk geometries
and thereby determines the random-matrix universality class.  In the
horizon matrix model the connection is microscopic: the antiunitary
symmetry acts directly on the fuzzy-sphere quantum geometry through
\eqref{eq:TR-q}, and the Dyson class of a conditioned parton
Hamiltonian is determined by whether the corresponding geometry is
invariant under this transformation.  Thus, in both cases,
antiunitary symmetry, quantum geometry, and random-matrix universality
are directly linked, with the present model providing a microscopic
realization of this connection for black-hole horizon dynamics.

\section{Discussion and Outlook}
\label{sec:discussion}

We have identified a microscopic route from quantum horizon geometry
to random-matrix spectral universality. On the round fuzzy sphere the
one-parton Hamiltonian consists of two exactly degenerate bands, the
same highly degenerate lowest-Landau-level states whose multiplicity
accounts for the Bekenstein--Hawking entropy. Physical $K=L\pm1$
fluctuations of the fuzzy sphere lift this degeneracy. Rotational
symmetry fixes the projected operator structure, a Wigner $9j$ symbol
determines the exact reduced couplings, and the finite-$N$ one-loop
stability spectrum fixes the vacuum widths. The resulting ensemble of
conditional Hamiltonians is therefore derived from the microscopic
horizon theory rather than postulated. Its spectral statistics agree
with the symmetry-selected Dyson classes at the level of adjacent-gap
ratios ($\chi^2=10.71$ for 8 degrees of freedom against matched GUE),
full spacing distributions ($D_{\rm KS}\le2.34\times10^{-3}$), and the
connected spectral form factor. The ensemble also realizes the
Pandey--Mehta GOE-to-GUE crossover in a microscopically determined
scaling variable $x=\eta\rho\sqrt N$. We
now summarize what these results teach us about the black hole
described by the model.

Entropy and spectral chaos originate in the same microscopic
parton sector.  The same set of microstates which
account for the black-hole entropy
also underlie the spectral chaos found here.
In this model, the entropy of the horizon is carried by the
multiplicity of the many-parton occupation states, while its spectral
chaos arises from the way quantum geometry lifts and mixes the
underlying one-parton degeneracy.

The randomness is quantum-geometric rather than thermal. Although
the horizon has the Hawking temperature, the geometric modes satisfy
$\Xi_{LK}=\hbar\omega_{LK}/T_H\gg1$, with $\Xi_{LK}=O(\sqrt N)$ or
larger for all non-exceptional modes and $\Xi_{10}\simeq52.6$ for the
scale mode, so thermal corrections to the geometric distribution are
$O(e^{-\Xi})$. The ensemble is therefore the Born distribution of the
geometric ground state.
 
Universal fine spectral correlations become largely insensitive
to microscopic geometric detail.
The covariance of the ensemble retains rank-dependent
weights and the full fuzzy-sphere operator basis, and is not invariant
in matrix space. Nevertheless, universal fine spectral
correlations become largely insensitive to microscopic geometric detail.
Once the symmetry class, finite dimension, and crossover coordinate are fixed,
the local spectral correlations tested here agree with the corresponding
RMT predictions. This parallels the gravitational situation, in which
the spectral form factor of a black hole displays the ramp and plateau
of random-matrix theory~\cite{Cotler:2016fpe} and semiclassical
gravity appears to compute ensemble-averaged spectral
quantities~\cite{Saad:2018bqo,Saad:2019lba}. 

The one-parton horizon Hamiltonian is highly connected and
  effectively nonlocal.
Fast scrambling,
$t_{\rm scr}\sim\beta\log S$, is believed to require nonlocal
interactions~\cite{Sekino:2008he,Lashkari:2011yi}. The rank-resolved
variance of Section \ref{operator_support},
computed with the finite-$N$ curvatures, shows
that the horizon Hamiltonian meets this requirement in a specific
geometric way. In SYK the all-to-all couplings are imposed.
Here they originate in the noncommutativity of the horizon geometry
and the same tensor operators enter the induced four-parton
couplings \eq{J4}.

Quantum geometry directly selects the chaos universality class.
The Dyson universality class governing the spectral chaos of a given
quantum-geometric realization of the horizon---GUE or GOE/GSE---is
fixed by the geometry itself, rather than assigned independently to
the spectrum. This
closely parallels the Stanford--Witten picture in JT gravity, where
antiunitary symmetry is encoded geometrically in the gravitational
path integral through the orientation structure of the bulk. In both
settings, quantum geometry is therefore not merely a passive
background for the chaotic dynamics. It determines whether the
relevant antiunitary symmetry is preserved or broken and thereby
selects the universality class governing the black-hole spectral
statistics.

The scope of these results should be stated precisely. They concern
the conditional one-parton Hamiltonian, whose spectrum contains $O(N)$
levels. The random-matrix predictions most closely associated with
black holes, in particular a plateau at times of order $e^{S}$,
concern the many-body spectrum. Non-interacting fermions with chaotic
single-particle levels do not produce Wigner--Dyson many-body
statistics, so the interaction term of the kind \eq{J4} is essential for this
step. Moreover, one-body random-matrix statistics are common to many
chaotic quantum systems. They constitute a necessary consistency
requirement for a microscopic horizon, not a signature unique to black
holes. No Lyapunov exponent has been computed, so neither the chaos
bound $\lambda_L\le2\pi/\beta$~\cite{Maldacena:2015waa} nor fast
scrambling has yet been tested.

Looking ahead, our results show that the microscopic horizon possesses
the ingredients expected of black-hole chaos: nonlocality,
noncommutativity, quantum-generated mixing of its entropy-carrying
states. The next step is to consider the many-body system
directly and work out diagnostics such as
the level statistics of the interacting Hamiltonian
with the factorized couplings $J_{ij;kl}(\o)=\sum_A
G_A(\o) (T_A)_{ij}(T_A)_{kl}$, and the OTOC, to extract
$\lambda_L$ and to determine whether the scrambling time $t_{\rm scr}/\beta$
scales as $\log
N^2$. 
Beyond these, the real-time evolution of $H_p[\bm q(t)]$,
including level-velocity statistics, transition and dephasing rates
and possible Dyson Brownian dynamics, together with a microscopic
derivation of the decoherence rate from the link sector, would connect
the spectral chaos found here to relaxation, dissipation and transport
on the horizon, and hence to the membrane response derived previously
in the same framework.

Taken as a whole, the horizon matrix model provides a remarkably
economical microscopic framework for black-hole physics.  Starting
from a single matrix Hamiltonian and its fuzzy-sphere background,
the model accounts for the Bekenstein--Hawking entropy at
large $N$ through a set of horizon degrees of freedom,
identifies the fermionic horizon
partons and their Landau-level structure, gives a microscopic
realization of membrane response, provides a tunneling description of
Hawking emission, exhibits quantum stabilization of the fuzzy-sphere
background, and, as shown here, generates the random-matrix spectral
statistics and highly connected dynamics expected of a chaotic black
hole.  These phenomena are not introduced through separate effective
models. They arise from different sectors and limits of the same
microscopic matrix dynamics.  This coherence is arguably the main
strength of the construction.

At the same time, an important limitation should be kept in view.
Gravity itself has so far entered only indirectly.  The model
describes the microscopic quantum dynamics of the horizon, while the
surrounding spacetime geometry, gravitational redshift, bulk
propagation, and Einstein dynamics remain outside the present microscopic
description, although consistency allows the quantum pressure of the
radial fuzzy-sphere mode to be balanced by a de Sitter interior,
suggesting a possible first connection to the bulk geometry.

The model is therefore, at present, most naturally regarded as a
microscopic model of the quantum horizon rather than as a complete
theory of quantum gravity.  Establishing a consistent coupling to
gravity, whether emergent from the matrix model itself or supplied by a
larger framework, will be essential for identifying this microscopic
system unambiguously with a black-hole horizon.
The central open question is to determine how gravity emerges from, and
couples to, the matrix degrees of freedom.


\appendix

\section{Projected perturbed Yukawa operator}
\label{app:WE}

Under a perturbation
\be \label{pert-Xa}
\delta X_a
=\sum_{LKQ} q_{LKQ}B_a^{LKQ}
\ee
of the fuzzy sphere, the
energy levels of the partons are obtained by diagonalizing the fermion
Hamiltonian
\be
h_F = g_N : \psi^\dag K(q) \psi:
\ee
where $K(q) :=\s_a X^a = K_0 + \d K(q)$ with $K_0 = \s_a J^a$ and
\be
\delta K(q)=\sum_{\rm phys.\ }q_{LKQ}\cO_{LKQ},\qquad
\cO_{LKQ}:=\sigma^aB_a^{LKQ}.
\ee
Within each fixed fermion band $k$,
it is therefore necessary to understand the structure of the operator
$P_k \cO_{LKQ} P_k$. We claim that it obeys a Wigner--Eckart theorem of the form:
\be \label{POP1}
P_k \cO_{LKQ} P_k = \Rp_{LK}^{(k)}\Tens_{KQ}^{(k)},
\ee
where the  tensor matrix $\Tens_{KQ}^{(k)}$ has the matrix elements 
\be\label{Tmatrix1}
\langle k,m'|\Tens_{KQ}^{(k)}|k,m\rangle
=(-1)^{k-m'}\sqrt{2K+1}
\begin{pmatrix}
k&K&k\\
-m'&Q&m
\end{pmatrix}
\ee
and is  Hilbert-Schmidt  normalized in the sense
\be
\Tr\!\left[(\Tens_{KQ}^{(k)})^\dagger\Tens_{K'Q'}^{(k)}\right]
=\delta_{KK'}\delta_{QQ'}.
\ee
The reduced coefficient $\Rp_{LK}^{(k)}$ is given in terms of the Wigner
9$j$-symbol as
\be \label{R9japp}
\Rp_{LK}^{(k)}
=(2k+1)\sqrt{6(2L+1)}
\begin{Bmatrix}
J&\frac12&k\\
J&\frac12&k\\
L&1&K
\end{Bmatrix}.
\ee

To prove this, let us substitute \eq{Ba} into  \eq{yuk}, we obtain
\be \label{OLKQ}
\mathcal O_{LKQ}
=\sqrt{2} \sum_{M,r}\langle LM;1r|KQ\rangle
T_{LM}\otimes\t_r.
\ee
Here
\be
\t_r:=\bm e_r\!\cdot\!\frac{\bm \sigma}{\sqrt{2}},
\quad r=0,\pm1, \qquad
\operatorname{tr}_{1/2}\!\left(\t_r^\dagger\t_{r'}\right)
=\delta_{rr'}
\ee
is a spherical tensor of rank 1 and is unit-normalized. 
Group theoretically,  $\mathcal O_{LKQ}$ is a spherical tensor of rank $K$
obtained from the multiplication of spherical tensors
of rank $L$ and rank $1$
\be
\mathcal O_{LKQ}
=\sqrt2\,[T^{(L)}\otimes\tau^{(1)}]_{KQ}.
\label{eq:OLKQtensorproduct}
\ee
In fact, it is easy to show that
\be
   [\cK_z, \cO_{LKQ}] = Q \cO_{LKQ}, \quad
   [\cK_\pm, \cO_{LKQ}] = \sqrt{(K\mp Q)(K\pm Q+1)} \,  \cO_{LK,Q\pm1},
   \ee
   therefore $\cO_{LKQ}$ is an irreducible rank-$K$ tensor operator on 
   $\mathcal H_J\otimes\mathcal H_{1/2} = \cH_{k_+} \oplus \cH_{k_-}$.
   We note that
   this tensor character survives projection to a fixed fermion band.
Indeed, $P_k$ projects onto an
irreducible eigenspace of $\cK^2$
and therefore commutes with the total rotations,
\be
[P_k, \cK_a] =0.
\ee
Consequently $P_k\mathcal O_{LKQ}P_k$ obeys the same rank-$K$
commutation relations as $\mathcal O_{LKQ}$ and hence it is itself
a rank-$K$ irreducible tensor operator acting on $\cH_k$.
The Wigner--Eckart theorem then fixes all of its $m,m',Q$
dependence and leaves only one reduced coefficient, which depends only on
$L, K$ and $k$.  This is basically the form
\eq{POP1}.

To explain better the meaning of the terms appearing in the relation \eq{POP1},
let us note that for a single spin-$k$ irreducible space,
\be
\operatorname{End}(\mathcal H_k)
\simeq \mathcal H_k\otimes\mathcal H_k^*
\simeq\bigoplus_{K'=0}^{2k}\mathcal H_{K'},
\label{eq:Enddecomp}
\ee
and every rank $K'$ occurs exactly once.
Hence, for $K\leq 2k$, there is, up to an overall normalization, a unique
rank-$K$ tensor operator in $\operatorname{End}(\mathcal H_k)$. Let
$\Tens_{KQ}^{(k)}$ be the Hilbert--Schmidt 
normalized representative, we finally obtain
\be
P_k\mathcal O_{LKQ}P_k
=\Rp_{LK}^{(k)}\Tens_{KQ}^{(k)},
\label{eq:WEstructural}
\ee
with a coefficient $\Rp_{LK}^{(k)}$ independent of $Q$. Note that by
definition, the unique Hilbert--Schmidt 
normalized rank $K$-tensor $\Tens_{KQ}^{(k)}$ on $\cH_k$ has the
matrix elements \eq{Tmatrix1}.

To derive the recoupling coefficient $\Rp_{LK}^{(k)}$ and explain how
the 9$j$ symbol arises, we utilize the  tensor-product Wigner--Eckart theorem,
which states that \cite{zare1988angular,Varshalovich:1988ifq}: 
For an operator $[T^{(k_1)}\otimes U^{(k_2)}]^{(K)}$ acting separately
on the two factors of a coupled state $(j_1j_2)j$,
the reduced matrix element of  $[T^{(k_1)}\otimes U^{(k_2)}]^{(K)}$ can be
expressed in terms of a Wigner 9$j$ symbol
\begin{align}
&\left\langle (j_1j_2)j\left\|
[T^{(k_1)}\otimes U^{(k_2)}]^{(K)}
\right\|(j_1'j_2')j'\right\rangle
\nonumber\\
&\quad=
\sqrt{(2j+1)(2j'+1)(2K+1)}
\begin{Bmatrix}
 j_1&j_2&j\\
 j_1'&j_2'&j'\\
 k_1&k_2&K
\end{Bmatrix}
\langle j_1\|T^{(k_1)}\|j_1'\rangle
\langle j_2\|U^{(k_2)}\|j_2'\rangle .
\label{eq:producttensorWE}
\end{align}
The $9j$ symbol therefore has a simple interpretation:
it is the reduced recoupling coefficient relating two different
angular-momentum coupling schemes -- one in which  the
subsystem angular momenta $j_1$ and $j_2$ are first combined to form
the total states, and one in which the tensor operators
$T^{(k_1)}$ and $U^{(k_2)}$ are first coupled to form a tensor
operator of total rank $K$.

For the present problem one simply substitutes
\be
 j_1=j_1'=J,
\qquad
 j_2=j_2'=\frac12,
\qquad
 j=j'=k,
\qquad
 k_1=L,
\qquad
 k_2=1.
\label{eq:producttensorsubstitution}
\ee
Equation~\eqref{eq:producttensorWE} then becomes
\be
\left\langle k\left\|\mathcal O_{LK}\right\|k\right\rangle
={}(2k+1)\sqrt{2(2K+1)}
\begin{Bmatrix}
J&\frac12&k\\
J&\frac12&k\\
L&1&K
\end{Bmatrix}
\langle J\|T^{(L)}\|J\rangle
\left\langle\frac12\left\|\tau^{(1)}\right\|\frac12\right\rangle .
\label{eq:Ored9jpre}
\ee
The two reduced matrix elements can be fixed by the
Hilbert--Schmidt normalization and the results are (see below)
\be \label{rrr}
\langle J\|T^{(L)}\|J\rangle = \sqrt{2L+1}, \qquad
\left\langle\frac12\left\|\tau^{(1)}\right\|\frac12\right\rangle = \sqrt{3}.
\ee
Consequently,
\be
\left\langle k\left\|\mathcal O_{LK}\right\|k\right\rangle
=(2k+1)\sqrt{6(2L+1)(2K+1)}
\begin{Bmatrix}
J&\frac12&k\\
J&\frac12&k\\
L&1&K
\end{Bmatrix}.
\label{eq:Ored9j}
\ee
Now, take the reduced matrix elements of \eq{POP1} and noticing that
\be
\langle k\|\Tens_K^{(k)}\|k\rangle=\sqrt{2K+1},
\label{eq:Tkredcompact}
\ee
which follows from its definition \eq{Tmatrix1}, we obtain
immediately \eq{R9japp}. 

Finally, let us establish \eq{rrr}. 
For the  scalar fuzzy harmonics $T_{LM}$.
Inside the spin-$J$ matrix space, the Wigner--Eckart theorem
gives 
\be
\langle J\mu'|T_{LM}|J\mu\rangle
=
(-1)^{J-\mu'}
\begin{pmatrix}
J&L&J\\
-\mu'&M&\mu
\end{pmatrix} \langle J\|T^{(L)}\|J\rangle.
\label{eq:TLMpre}
\ee
To determine the reduced matrix element
$\langle J\|T^{(L)}\|J\rangle$, we insert \eqref{eq:TLMpre} into
the normalization condition for  $T_{LM}$:
\be
\Tr\!\left(T_{LM}^{\dagger}T_{L'M'}\right)=\delta_{LL'}\delta_{MM'}
\label{eq:TLMHS}
\ee
and utilize the $3j$ orthogonality relation
\be
\sum_{\mu',\mu}
\begin{pmatrix}
J&L&J\\
-\mu'&M&\mu
\end{pmatrix}^{\!2}
=\frac{1}{2L+1},
\label{eq:3jorthL}
\ee
we get $\langle J\|T^{(L)}\|J\rangle= \sqrt{2L+1}$.
Using the standard angular-momentum sign convention,
we finally obtain
\be
\langle J\mu'|T_{LM}|J\mu\rangle
=(-1)^{J-\mu'}\sqrt{2L+1}
\begin{pmatrix}
J&L&J\\
-\mu'&M&\mu
\end{pmatrix} .
\label{eq:TLMmatrix}
\ee
Similarly, for the normalized spin-one tensor $\t_r$,
we obtain
$\langle \frac{1}{2}\|\tau^{(1)}\|\frac{1}{2}\rangle= \sqrt{3}$,  and so
\be
\langle \tfrac12\alpha'|\sigma_r|\tfrac12\alpha\rangle
=(-1)^{\frac12-\alpha'}\sqrt6
\begin{pmatrix}
\frac12&1&\frac12\\
-\alpha'&r&\alpha
\end{pmatrix}.
\label{eq:sigmamatrix}
\ee

\section{Numerical construction and spectral analysis}
\label{app:numerics}

In this appendix, we give details of the numerical construction used in
Sec.~\ref{sec:RMT}.  The calculation proceeds directly from the
finite-$N$ quantum-corrected bosonic curvatures and the exact projected
tensor operators.  No random-matrix distribution is imposed on the
microscopic Hamiltonian.  Rather, independent vacuum coordinates are
sampled from the geometric Born distribution, the corresponding
conditional parton Hamiltonians are constructed and diagonalized, and
their spectral observables are subsequently compared with matched
finite-dimensional random-matrix ensembles.

\subsection{Branch-resolved construction of the Hamiltonian}

For a fixed rank $N$, the two parton bands have
\be
 k_u=\frac{N}{2},\quad d_u=N+1,
 \qquad\qquad
 k_d=\frac{N-2}{2},\quad d_d=N-1 .
 \label{eq:app-banddim}
\ee
For every physical bosonic branch $(L,K)$ with $K=L\pm1$, we use the
positive finite-$N$ curvature
$\Lambda^{\rm eff}_{LK}(N)$ obtained from the one-loop stability
calculation \cite{Chu:2026dhx}.
The gauge branch $K=L$ is absent.  The $K=0$ mode is also
omitted from the spectral analysis, since its projection gives a common
intraband energy shift and therefore does not split the parton levels.

It is convenient to define the scale-free branch coefficient
\be
 a^{(s)}_{LK}
 :=
 \frac{\cR^{(s)}_{LK}}
      {[\Lambda^{\rm eff}_{LK}(N)]^{1/4}},
 \label{eq:app-branchweight}
\ee
where $\cR^{(s)}_{LK}$ denotes the exact finite-$N$ reduced coefficient
given in Eqs.~\eq{R+}, \eq{R-} for the corresponding
parton band.  Thus the scale-free conditional Hamiltonian can be written
as
\be
 \cH_s(\bm\xi)
 =
 \sum_{L,K=L\pm1}\sum_A
 a^{(s)}_{LK}\,
 \xi_{LK,A}\,
 \Mten^{(k_s)}_{K,A}.
 \label{eq:app-Hnumeric}
\ee
Here
\be
 \xi_{LK,A}\sim{\cal N}(0,1)
\ee
are independent real Gaussian variables for every physical branch and
every one of the $2K+1$ Hermitian tensor components
\be
 A=\{0,c1,s1,\ldots,cK,sK\}.
\ee
An important point is that the same realization
$\{\xi_{LK,A}\}$ is used to construct both the $u$ and $d$ Hamiltonians:
the two parton bands respond to the same fluctuating horizon geometry.

The tensor matrices are constructed from
\be
 \langle k,m'|T^{(k)}_{KQ}|k,m\rangle
 =
 (-1)^{k-m'}\sqrt{2K+1}
 \begin{pmatrix}
  k&K&k\\
  -m'&Q&m
 \end{pmatrix},
 \label{eq:app-Tmatrix}
\ee
and are combined into the real Hermitian basis of
Eq.~\eqref{real-b}.  The matrices are Hilbert--Schmidt
orthonormal.  For the $d$ band, the minus transpose associated with the
hole description is incorporated in $\Mten^{(k_d)}_{K,A}$ as in
Eq.~\eqref{Mten}.
The constant round-sphere energies and the common multiplicative factor
\be
 \Gamma_N=\frac{g_N\sqrt{a_0N}}{2}
\ee
are omitted.  Neither affects the spectral statistics, while the overall
spectral scale is removed by unfolding for the observables that depend
on unfolded energies.  Each resulting Hermitian matrix is diagonalized
and its eigenvalues are ordered before the spectral statistics are
computed.

\subsection{Symmetry sectors and numerical validation}

Two ensembles are constructed from every geometric realization.  The
full physical Hamiltonian contains all $K>0$ tensor ranks, whereas the
time-reversal-invariant ensemble retains only even $K$,
\be
 \cH_{s,e}
 =
 \sum_{K\ {\rm even}} \cH^{(s)}_K .
\ee
This provides a direct numerical implementation of the symmetry
classification derived in Sec.~\ref{sec:symmetry}.

Several internal checks are performed before evaluating the spectral
statistics.  The closed expressions \eq{R+}, \eq{R-}
for $\cR^{(s)}_{LK}$ are compared
with a direct evaluation of the corresponding Wigner-$9j$
formula
\be
 \cR^{(k)}_{LK}
 =
 (2k+1)\sqrt{6(2L+1)}
 \begin{Bmatrix}
 J&\frac12&k\\
 J&\frac12&k\\
 L&1&K
 \end{Bmatrix}.
\ee
The tensor matrices are spot-checked against their explicit
$3j$ representation and their Hermiticity and normalization are verified
numerically.  Finally, the assembled Hamiltonians are checked to be
Hermitian within numerical precision.

Time-reversal invariance of the even-$K$ Hamiltonian is checked
directly.  Writing
\be
 \Theta=U{\cal K},
\ee
where ${\cal K}$ denotes complex conjugation, we monitor the relative
residual
\be
 \epsilon_{\rm TR}
 =
 \frac{\|\cH_{s,e}-U s\cH_{s,e}^*U^\dagger\|_F}
      {\|\cH_{s,e}\|_F}
 \label{eq:app-TRres}
\ee
using the Frobenius norm $\|  A \|_F := \sqrt{\Tr A^\dag A}$. 
For odd $N$, $\Theta^2=-1$ and the even-$K$ spectra consist of Kramers
pairs.  After diagonalization, the ordered eigenvalues are grouped as
\be
 (E_1,E_2),\;(E_3,E_4),\ldots .
\ee
For each pair we verify that the relative splitting is negligible
within numerical precision.  The two eigenvalues are then replaced by
their mean, and all GSE spectral statistics are computed from the
resulting set of distinct Kramers-reduced levels. 

\subsection{Adjacent-gap ratio and matched finite-dimensional ensembles}

The adjacent-gap ratio is evaluated directly from the raw ordered
spectrum,
\be
 \delta_n=E_{n+1}-E_n,\qquad
 r_n=
 \frac{\min(\delta_n,\delta_{n+1})}
      {\max(\delta_n,\delta_{n+1})},
 \qquad
 \bar r=\frac{1}{d-2}\sum_{n=1}^{d-2}r_n .
 \label{eq:app-r}
\ee
No unfolding is required.  The two bands are always analyzed
separately.  For a given realization their estimates are combined only
afterward according to the number of available gap ratios,
\be
 \bar r_{\rm pool}
 =
 \frac{(d_u-2)\bar r_u+(d_d-2)\bar r_d}
      {(d_u-2)+(d_d-2)} .
 \label{eq:app-rpool}
\ee
In the GSE sector, $d_u$ and $d_d$ in this expression denote the
Kramers-reduced dimensions.
The $u$ and $d$ bands are treated as distinct spectral sectors.
All level statistics are computed within each band before the
band-resolved results are combined with the appropriate statistical
weights; the two eigenvalue spectra are never merged into a single
ordered sequence.

For every microscopic data set, we generate a corresponding
finite-dimensional RMT reference ensemble with the corresponding retained
dimension and symmetry class.  A GOE matrix is obtained by symmetrizing
a real Gaussian matrix,
\be
 H_{\rm GOE}=\frac{A+A^T}{2},
\ee
while a GUE matrix is obtained by Hermitizing a complex Gaussian matrix,
\be
 H_{\rm GUE}=\frac{A+A^\dagger}{2}.
\ee
For the GSE reference ensemble, the matrix dimension is even,
$d=2m$.  We introduce
\be
 U=
 \begin{pmatrix}
 0&I_m\\
 -I_m&0
 \end{pmatrix},
 \qquad
 \Theta=U{\cal K},
 \qquad
 \Theta^2=-1 .
\ee
Starting from a generic complex Hermitian Gaussian matrix $H_0$, we
construct
\be
 H_{\rm GSE}
 =
 \frac12
 \left(
 H_0+UH_0^*U^\dagger
 \right).
\ee
By construction,
\be
 UH_{\rm GSE}^*U^\dagger=H_{\rm GSE},
\ee
so that $H_{\rm GSE}$ is time-reversal invariant with
$\Theta^2=-1$.  Its eigenvalues therefore occur in Kramers-degenerate
pairs.  After verifying that the numerical splitting within each pair is
negligible, each pair is replaced by its mean and the spectral
statistics are computed from the resulting set of distinct levels.
The overall variance normalization of these
reference matrices is immaterial for $\bar r$.

The production calculation uses $12,000$ independent geometric
realizations at
\be
N=12,15,20,25,30,45,50,60 .
\ee
For each $N$, the same set of geometric realizations is used to
construct both the full physical Hamiltonian and its even-$K$
time-reversal-invariant restriction.  Thus the full-ensemble and
even-$K$ results reported in Tables~2 and 3 are obtained from a
single common microscopic production sample.
For the adjacent-gap-ratio comparison, we generate $12,000$
independent realizations of each corresponding matched
finite-dimensional RMT ensemble.
Likewise, the matched GUE entries appearing in
the two tables are obtained from the same GUE reference sample.

\subsection{Ensemble unfolding}

Random-matrix universality concerns correlations among the fluctuating
energy levels rather than the smooth, system-dependent variation of the
mean spectral density.  Before comparing the microscopic spectra with
RMT, one must therefore remove this non-universal smooth density while
leaving the spectral fluctuations themselves unchanged.  This is achieved by
unfolding the spectrum \cite{Mehta2004RandomMatrices}.
For a realization $a$ with ordered levels $E_{a,n}$, the spectral
staircase is
\be
 {\cal N}_a(E)
 =
 \sum_{n=1}^{d}\Theta(E-E_{a,n}),
\ee
which counts the number of levels below $E$.  Its smooth
ensemble-averaged part, denoted by $\overline{\cal N}(E)$, determines
the mean spectral density through
\be
 \overline{\rho}(E)
 =
 \frac{d\overline{\cal N}(E)}{dE}.
\ee
The unfolded levels are then defined by
\be
 \widetilde E
 =
 \overline{\cal N}(E).
\ee
Consequently, $  d\widetilde E  = \overline{\rho}(E) dE$
and the unfolded spectrum has unit mean level density.  Spectral
correlations can therefore be compared directly between the microscopic
ensemble and the corresponding finite-dimensional RMT ensemble.

To construct the unfolding map $\overline{\cal N}(E)$ numerically, we define,
after any required symmetry is resolved, the ensemble-averaged
ordered spectrum
\be
 \mu_n:=\langle E_{a,n}\rangle_a,
 \qquad n=1,\ldots,d .
\ee
The smooth mean staircase is then determined from the relation
\be
 \overline{\cal N}(\mu_n)\simeq n .
\ee
Rather than fitting each realization separately, a single smooth
function is obtained from the ensemble-averaged spectrum and applied
identically to all realizations.  This removes the common smooth
variation of the density without fitting away realization-dependent
spectral fluctuations.

For numerical convenience, the mean spectrum is centered and rescaled
to the interval $[-1,1]$ by introducing
\be\label{eq:app-u}
 e_c:=\frac{\mu_1+\mu_d}{2},
 \qquad
 e_s:=\frac{\mu_d-\mu_1}{2},
 \qquad
 u_n:=\frac{\mu_n-e_c}{e_s}.
\ee
Thus $u_1=-1$ and $u_d=1$.
This rescaling is not part of the unfolding itself; it avoids large
differences in scale among the polynomial terms and improves the
numerical robustness of the fit.
Since the ensemble-averaged spectrum is symmetric about its center, the
centered staircase is fitted by the low-order odd polynomial
\be\label{eq:app-unfold-fit}
n-\frac{d+1}{2}
\approx
F(u_n),
\qquad
F(u)=a_1u+a_3u^3+a_5u^5
\ee
using least squares, 
the unfolding map is therefore
\be\label{eq:app-unfold-map}
 \overline{\cal N}(E)
 =
 \frac{d+1}{2}
 +
 F\!\left(\frac{E-e_c}{e_s}\right).
\ee
The cubic and quintic terms allow the smooth finite-dimensional
variation of the mean density to be captured without introducing a
high-order fit that could follow short-scale spectral structure.

The same map is finally applied to every level in every realization,
\be
 \widetilde E_{a,n}
 =
 \overline{\cal N}(E_{a,n}) .
\ee
For levels lying outside the range of the mean spectrum, the map is
continued linearly using the tangent at the corresponding boundary,
rather than by polynomial extrapolation.
Since the smooth fit gives unit mean spacing only approximately, we
apply one final common rescaling to all realizations,
\be
 x_{a,n}
 :=
 \frac{\widetilde E_{a,n}}{\bar s},
 \qquad
 \bar s
 =
 \left\langle
 \widetilde E_{a,n+1}-\widetilde E_{a,n}
 \right\rangle_{a,n},
\ee
so that the ensemble-averaged unfolded spacing is exactly unity.
No additional normalization is applied to individual realizations.

\subsection{Spacing distribution and longer-range observables}

With the unfolded levels defined by $x_{a,n}$, 
one can introduce the unfolded nearest-neighbor spacings 
\be
 s_{a,n}=x_{a,n+1}-x_{a,n},
 \qquad
 \langle s\rangle=1 .
\ee
Spacing samples are first constructed separately for the two bands.
They may then be pooled at the level of the spacing data,
 while the
underlying eigenvalue spectra remain separate.  The resulting pooled
spacing sample is used to construct the nearest-neighbor spacing
distribution $P(s)$.

The agreement of the microscopic and RMT spacing distributions is
quantified using the two-sample Kolmogorov--Smirnov test:
\be
 D_{\rm KS}
 :=
 \sup_s
 \left|
 F_{\rm hor}(s)-F_{\rm RMT}(s)
 \right| ,
\ee
where 
\be
F(s) = \frac{1}{M}\sum_{i=1}^{M} \Theta(s-s_i),
\ee
denotes the empirical cumulative distribution function (CDF)
constructed from the sample $\{ s_i \}$, and
$F_{\rm hor}(s)$ and $F_{\rm RMT}(s)$ denote the CDF for
the microscopic horizon and matched RMT spacing samples, respectively.
The Wigner surmises are quoted only as analytic guides to the expected
spacing distributions and level-repulsion exponents.
All quantitative
comparisons are instead made with the matched finite-dimensional RMT
ensembles of the same matrix dimension and symmetry class.

For each realization, the spectral trace is
\be
 Z_a(\tau)
 =
 \sum_{n=1}^{d}
 e^{-2\pi i\tau x_{a,n}} .
\ee
The full and connected spectral form factors are
\be
 K(\tau)
 =
 \frac1d
 \left\langle |Z_a(\tau)|^2\right\rangle_a, \qquad 
 K_c(\tau)
 =
 \frac1d
 \left[
 \left\langle |Z_a(\tau)|^2\right\rangle_a
 -
 \left|\left\langle Z_a(\tau)\right\rangle_a\right|^2
 \right].
 \label{eq:app-SFF}
 \ee
 The spectral form factors are evaluated on the full grid
$0\le\tau\le2$. For the quantitative comparison, however, the
connected-form-factor RMSE
\be
 {\rm RMSE}_c
 =
 \left[
 \frac1{N_\tau}
 \sum_{\tau_i}
 \left(
 K_c^{\rm hor}(\tau_i)
 -
 K_c^{\rm RMT}(\tau_i)
 \right)^2
 \right]^{1/2},
 \label{eq:app-SFFrmse}
\ee
is evaluated over the restricted interval
$0.05\le\tau\le1.50$, in order to exclude the short-time region most
sensitive to nonuniversal finite-size effects and to avoid overweighting
the late-time plateau, where agreement is comparatively trivial.
The comparison reported in the text is evaluated
at the 291 grid points
$\tau=0.05,0.055,\ldots,1.50$

\subsection{GOE-to-GUE crossover}

For even $N$, each microscopic realization is decomposed into
\be
 \cH_s
 =
 \cH_{s,e}+\cH_{s,o},
\ee
where $\cH_{s,e}$ contains the even-$K$ tensor components and $\cH_{s,o}$
contains the odd-$K$ components.  The interpolating Hamiltonian is
\be
 \cH_s(\eta)
 =
 \cH_{s,e}+\eta \cH_{s,o}.
 \label{eq:app-crossH}
\ee
The two pieces are obtained from the same underlying geometric
realization before the odd sector is multiplied by $\eta$.
The relative symmetry-breaking strength is evaluated directly from the
branch-resolved covariance.  With
\begin{align}
 \cV_{e}
 &:=
 \sum_s\sum_{K\ {\rm even}}
 (2K+1)w_{K,s}^2,
 \\
 \cV_{o}
 &:=
 \sum_s\sum_{K\ {\rm odd}}
 (2K+1)w_{K,s}^2,
\end{align}
we define
\be
 \rho
 =
 \sqrt{\frac{\cV_{o}}{\cV_{e}}}.
 \label{eq:app-rho}
\ee
Thus $\rho$ is computed before any spectral crossover fit is performed.
With this we introduce the variable
\be
 x :=\eta\rho\sqrt N,
\ee
and we propose to identify this with
the scaled Pandey--Mehta crossover parameter for the horizon model.

The matched finite-dimensional Pandey--Mehta reference is generated as
\be
 H_{\rm PM}(x)
 =
 S+i\frac{x}{\sqrt N}A,
 \label{eq:app-PM}
\ee
where $S$ is a real symmetric Gaussian matrix and $A$ is an independent
real antisymmetric Gaussian matrix.  Their independent off-diagonal
elements are normalized to have the same variance.  For every $x$ the
two dimensions $N+1$ and $N-1$ are diagonalized and analyzed separately,
and their adjacent-gap statistics are pooled using
Eq.~\eqref{eq:app-rpool}.

The numerical crossover scan is performed on the common scaled grid
\be
 x=0,0.05,0.10,\ldots,2.50 .
\ee
For each $N$, the corresponding microscopic deformation is
\be
 \eta=\frac{x}{\rho\sqrt N}.
\ee
We use $4000$ microscopic realizations for $N=20$ and $30$ and
$1000$ for $N=60$, together with equal numbers of matched
Pandey--Mehta realizations.
The crossover scale is therefore
fixed entirely by the microscopic quantity $\rho_N\sqrt N$. No
additional $N$-dependent normalization is introduced.



\section*{Acknowledgments}

We thank Norihiro Iizuka for discussion and comments.
We
acknowledge the support of this work by NCTS, the National Science and
Technology Council of Taiwan for the grant 113-2112-M-007-039-MY3, and
the National Tsing Hua University 2025 Talent Development Fund for a
TSAI WANG, YUAN-YANG Distinguished Talent Chair Professorship.
Numerical computations were carried out with assistance from OpenAI
ChatGPT and Codex. Implementation procedures and numerical 
results were independently checked by the author, who takes full
responsibility for the content of this work.

\bibliographystyle{utphys}

\bibliography{references}   
\end{document}